\documentclass[12pt,a4paper]{article}

\usepackage{graphicx}
\usepackage{xcolor}
\usepackage{siunitx}
\usepackage{amsmath, bm}
\usepackage{txfonts}
\usepackage[round]{natbib}
\usepackage[hidelinks]{hyperref}
\hypersetup{
    colorlinks,
    linkcolor={blue},
    citecolor={blue},
    urlcolor={blue}
}

\title{Can metal-rich worlds form by giant impacts?}

\date{%
Saverio Cambioni$^{1,*}$,
Benjamin P. Weiss$^{1}$, 
Erik Asphaug$^{2}$,\\
Kathryn Volk$^{2,3}$,
Alexandre Emsenhuber$^{4}$,
John B. Biersteker$^{1}$,
Zifan Lin$^{1}$,
Robert Melikyan$^{2}$~\\
~\\
\begin{small}
    $^1$Department of Earth, Atmospheric and Planetary Sciences, Massachusetts Institute of Technology, 77 Massachusetts Ave Bldg 54, Cambridge, MA 02139\\%
    ~\\
    $^2$Lunar and Planetary Laboratory, The University of Arizona, 1629 E University Blvd, Tucson, AZ 85721\\%
    ~\\
    $^3$Planetary Science Institute, 1700 E Fort Lowell Rd STE 106, Tucson, AZ 85719\\%
    ~\\
    $^4$Universitäts-Sternwarte, Ludwig-Maximilians-Universität München, Scheinerstraße 1, 81679 München, Germany\\
    ~\\
    *Corresponding author: cambioni@mit.edu
\end{small}
}

\begin{document} 

\maketitle

Planets and stars are expected to be compositionally linked because they accrete from the same material reservoir. However, astronomical observations revealed the existence of exoplanets whose bulk density is far higher than what is expected from host-stars’ composition. A commonly-invoked theory is that these high-density exoplanets are the metallic cores of super-Earth-sized planets whose rocky mantles were stripped by giant impacts. Here, by combining orbital dynamics and impact physics, we show that mantle-stripping giant impacts between super-Earths are unlikely to occur at rates sufficient to explain the observed size and currently estimated abundance of the high-density exoplanets. We explain this as the interplay of two main factors: the parent super-Earths being in most cases smaller than 2 Earth radii; and the efficiency of mantle stripping decreasing with increasing planetary size. We conclude that most of the observed high-density exoplanets are unlikely to be metal-rich giant-impact remnants.



\section{Introduction}

Planet formation models commonly assume that the abundance of refractory planet-forming elements in protoplanetary disks (e.g., Fe, Mg, Si) matches that inferred from spectroscopic observations of planet-hosting stars. However, the theory that planets and host stars are always compositionally linked is challenged by the existence of high-density exoplanets, here defined as cosmochemical outliers with bulk density higher than the 99.7th percentile of values expected from the host-star compositions \citep[Section \ref{sec:HD_exoplanets}]{unterborn2022nominal}. Exoplanetary survey data suggest that these high-density exoplanets may account for $\sim$9\% of all terrestrial planets (Section \ref{sec:observed_occurrence_rate}). How such a large population of high-density exoplanets formed is unknown, but two main theories exist.

High-density exoplanets could be the compressed silicate-iron cores of giant planets that lost their volatile-rich envelopes through tidal stripping during close encounters with their host stars or as the result of photoevaporation driven by extreme ultraviolet stellar radiation  \citep{2005IcarFaber,2014RSPTAMocquet}. In this case, the observed high-density exoplanets are predicted to remain high-density up to ${\sim}$1 billion years (Gyr) as the core decompresses following loss of the envelope \citep{2014RSPTAMocquet}. Alternatively, high-density exoplanets could be metal-rich worlds with a high (${\ge}~50\text{--}60$ wt.\%) iron content compared with Earth, whose metal content (${\sim}30$ wt.\%) is close to that of the CI carbonaceous chondrites and the Sun \citep{2003TrGePalme}. Examples of metal-rich worlds in the solar system are the planet Mercury and asteroid (16) Psyche \citep{2018mvambookEbel,2022SSRvElkins}.

In turn, metal-rich worlds could form in at least two ways. First, metal-rich worlds may form via preferential accretion of metal-rich materials near the condensation/sublimation line of refractory elements in the innermost part of protoplanetary disks \citep[e.g.,][]{2020ApJAguichine,2022AAJohansen}. Second, differentiated planets with a rocky mantle and a metallic core and with a metal-to-silicate ratio comparable to their host stars' may evolve into metal-rich worlds by losing their mantles in giant impacts \citep[e.g.,][]{2014NatGeoAsphaug}. Recent advances in analytical impact theory and modelling have enabled tests of the latter hypothesis. For example, previous studies have shown that Mercury analogs may form as a result of collisional grinding of terrestrial embryos in the early solar system \citep[e.g.,][]{2021PSJCambioni}, when the gravitational interactions between the growing planets overwhelm the damping effects of the dissipating nebular gas
and smaller planetesimals, leading to crossing orbits \citep{2006AJKenyonBromley,2006IcarusOBrien}.

\begin{figure*}
\centering
\includegraphics[width=\linewidth]{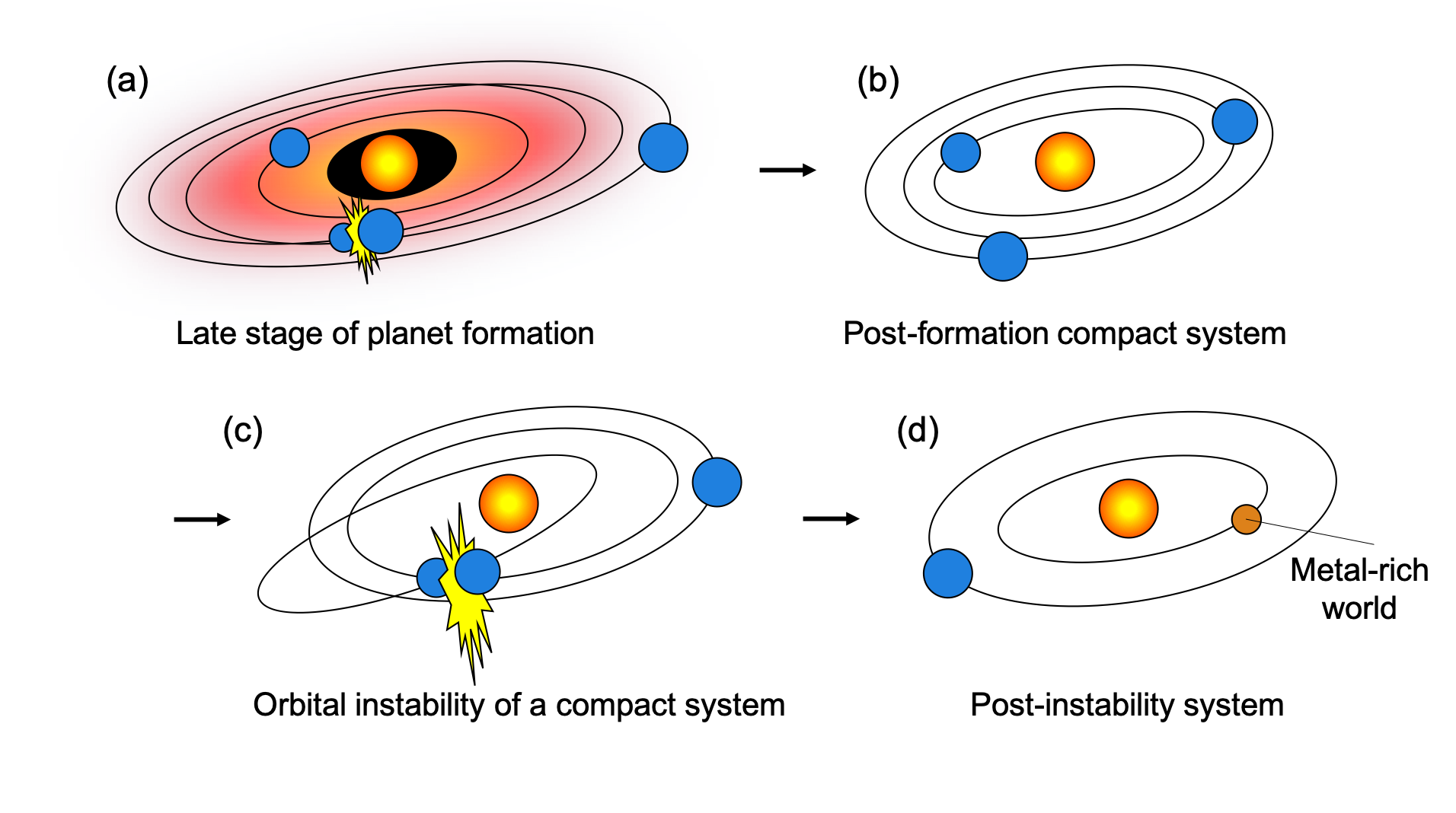}
\caption{\textbf{Are the observed high-density exoplanets metal-rich remnants of giant impacts?} \textbf{a}, Super-Earths are predicted to complete their formation through giant impacts whose energies are too low to strip their rocky mantles and form metal-rich planets \citep{2020MNRASScora,2022ApJScora,2020MNRASPoon,2022MNRASEsteves,2022AJGoldbergBatygin}. \textbf{b}, After their formation ceases, super-Earths may remain on closely spaced orbits typical of observed compact systems \citep{2022PPVIIWeiss} and can experience instabilities over timescales comparable to the main-sequence lifetime of the host stars \citep{2015ApJVolkGladman,2015ApJPuWu,2020PNASTamayo,2020AJVolk}. \textbf{c}, If a compact system becomes unstable, super-Earths may experience a second stage of giant impacts that are more energetic than those that occurred during their formation \citep{2015ApJVolkGladman}. These late giant impacts may erode the silicate mantles of the super-Earths to form metal-rich, high-density planets. \textbf{d}, The final, post-instability planetary system may have a metal-rich world with density akin to the measured densities of observed high-density exoplanets (Table \ref{tab:metal_rich_Pgi}). Colors indicate different planetary compositions: orange is metal-rich, blue indicates planets with a terrestrial composition (that is, differentiated in an iron core and a rocky mantle).}
\label{fig:cartoon_model}
\end{figure*}

Analogously to what proposed for planet Mercury, the observed high-density exoplanets could be the metallic cores of super-Earths that lost their mantles in giant impacts \citep{2009ApJMarcus,2022MNRASReinhardt}. However, because the gravitational binding energy of a planet increases with planetary size to the fifth power, stripping a super-Earth's mantle requires collisions with at least one order of magnitude higher specific impact energy than those needed to remove the mantle of proto-Mercury \citep{2009ApJMarcus,2022MNRASReinhardt,2023AREPSGabrielCambioni}. For a given colliding pair, the impact energy is defined by the collision velocity, $v_{coll}$,

\begin{equation}
\label{eq:v_coll}
v_{coll}= \sqrt{v_{esc}^2+v_\infty^2},
\end{equation}

\noindent
where $v_{esc}$ is the mutual escape velocity defined as

\begin{equation}
    v_{esc} = \sqrt{2G\frac{M_T+M_P}{R_T+R_P}},
\end{equation}

\noindent
where $G$ is the gravitational constant, $M$ is the planetary mass, $R$ is the planetary radius, and the subscripts ``$T$'' and ``$P$'' indicate the target and projectile of the collision, here defined as the bodies with the largest and smallest binding energies, respectively. The variable $v_{\infty}$ is the pre-encounter relative velocity of the colliding bodies. During the growth of super-Earths, $v_{\infty}$ is predicted to be small, leading to merging of the colliding bodies at $v_{coll}\sim v_{esc}$ \citep[Figure \ref{fig:cartoon_model}a;][]{2020MNRASScora,2020MNRASPoon,2022MNRASEsteves,2022AJGoldbergBatygin}. In studies that maximize the likelihood of mantle stripping  
by assuming that planetary embryos are small and on eccentric orbits \citep[that is, high $v_{\infty}$ and low $v_{esc}$, e.g.,][]{2022ApJScora},  the resulting metal-rich planets are smaller than GJ 367 b, which we identify as the smallest observed high-density exoplanet (Table \ref{tab:metal_rich_Pgi}; Section \ref{sec:HD_exoplanets}).

\begin{table}[t]
\begin{center}
\begin{tabular}{|l|c|c|c|c|} 

\hline
\textbf{Planet}	&	$\bm{R~(R_\oplus)}$			&	$\bm{M~(M_\oplus)}$			&	$\bm{\rho/\bar{\rho}}$ &	\textbf{$\bm{p_{HD}}$}\\
\hline
GJ 367 b	&	0.69	$\pm$	0.02	&	0.6	$\pm$	0.1	&	2.3 $\pm$	0.3	&	100	\\
K2-229 b	&	1.00	$\pm$	0.02	&	2.5	$\pm$	0.4	&	2.7	$\pm$	0.5	&	100 \\
GJ 3929 b	&	1.09	$\pm$	0.04	&	1.8	$\pm$	0.5	&	1.4	$\pm$	0.4	&	68	\\
TOI-1468 b	&	1.28	$\pm$	0.04	&	3.2	$\pm$	0.2	&	1.4	$\pm$	0.2	&	89	\\
TOI-431 b	&	1.28	$\pm$	0.04	&	3.1	$\pm$	0.4	&	1.4	$\pm$	0.2	&	77	\\
HD 137496 b	&	1.31	$\pm$	0.06	&	4.0	$\pm$	0.6	&	1.7	$\pm$	0.3	&	91	\\
Kepler-105 c	&	1.31	$\pm$	0.07	&	4.6	$\pm$	0.9	&	1.9	$\pm$	0.5	&	93	\\
HIP 29442 d & 1.37 $\pm$ 0.11 & 5.1 $\pm$ 0.4 & 1.8 $\pm$ 0.5 & 88\\
L 168-9 b	&	1.39	$\pm$	0.09	&	4.6	$\pm$	0.6	&	1.5	$\pm$	0.4	&	78	\\
Kepler-406 b	&	1.43	$\pm$	0.03	&	6.4	$\pm$	1.4	&	1.8	$\pm$	0.4	&	94	\\
Kepler-80 d	&	1.53	$\pm$	0.09	&	6.8	$\pm$	0.7	&	1.5	$\pm$	0.3	&	81	\\
Kepler-107 c	&	1.60	$\pm$	0.03	&	10.0	$\pm$	2.0	&	1.9	$\pm$	0.4	&	96\\
\hline
\end{tabular}
\end{center}
\caption{List of high-density exoplanets, their masses $M$, radii $R$, densities $\rho$ (in units of $\bar{\rho}$, see text) and probability of being high-density, $p_{HD}$ computed by means of Eq. \ref{eq:high-density_treshold}. We estimate the occurrence rate of high-density exoplanets among the terrestrial planets to be $\eta_{HD}\sim$ 9\% (Section \ref{sec:observed_occurrence_rate}). $\oplus$ = Earth's value.}
\label{tab:metal_rich_Pgi}
\end{table}

Overall, planet formation models predict that giant impacts between growing super-Earths generally occur at impact energies lower than those necessary for mantle stripping. After formation ceases, super-Earths may remain on short-period orbits that are tightly packed around their host stars; these configurations are commonly referred to as compact systems \citep[][Figure \ref{fig:cartoon_model}b]{2022PPVIIWeiss}. Planets in compact systems may experience eccentricity growth due to the overlapping of mean-motion resonances and secular resonances \citep{2015ApJVolkGladman,2015ApJPuWu,2020PNASTamayo,2020AJVolk}. This may eventually result in orbital instability (Figure \ref{fig:cartoon_model}c) and a phase of crossing orbits and giant impacts that occurs significantly after the initial era of planet formation [$>$10 million years (Myr), potentially up to Gyr \citep{2015ApJVolkGladman,2020PNASTamayo}]. Because $v_{\infty}$ increases with decreasing stellocentric distances and increasing mutual eccentricities \citep{2023AREPSGabrielCambioni}, giant impacts during this late period of orbital instability are likely to be more energetic, and thus more likely to strip the super-Earths' mantles, than impacts occurring in previous planet formation eras.

Motivated by this, here we combine exoplanetary observations, \textit{N}-body simulations, smoothed particle hydrodynamic (SPH) simulations of giant impacts, and machine learning to test the hypothesis that giant impacts in unstable compact systems of super-Earths can form metal-rich planets akin to the observed high-density exoplanets (Figure \ref{fig:cartoon_model}d). In particular, we set an upper limit on the rate with which orbital instabilities in compact systems can produce metal-rich giant-impact remnants.

\section{Data and methods}
\label{sec:data_methods}
\subsection{High-density exoplanets}
\label{sec:HD_exoplanets}

\begin{figure}[p]
	\centering	\includegraphics[width=0.95\linewidth]{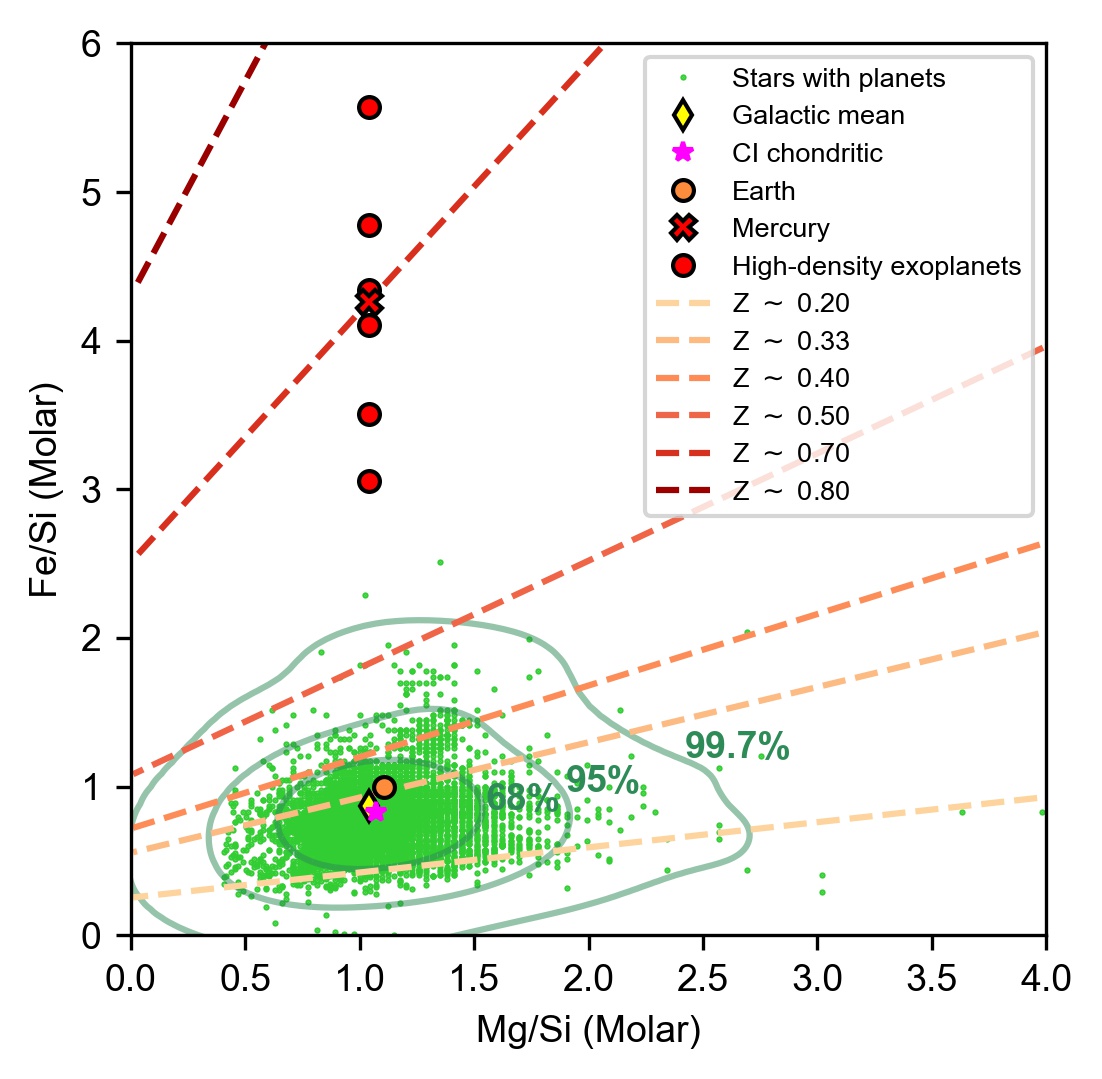}
	\caption{\textbf{Abundance of iron and rock-forming elements in the galactic neighborhood, selected planets and meteorites}. The green data points are the iron-to-silicon (Fe/Si) molar ratio versus magnesium-to-silicon (Mg/Si) molar ratio of stars with planets in the \href{https://www.hypatiacatalog.com/}{Hypatia} catalog by \citet{2014AJHinkel}, along with the 68\% ($1\sigma$), 95\% ($2\sigma$) and 99.7\% ($3\sigma$) isocontours and the (Fe/Si, Mg/Si) of the CI carbonaceous chondrites, Earth, Mercury and some of the high-density exoplanets in Table \ref{tab:metal_rich_Pgi}. Assuming a CI chondritic Mg/Si for the high-density exoplanets and Mercury, we derive their bulk Fe/Si from estimates of their core-mass fraction, $Z$, based on the measured mass and radii. We convert $Z$ into Fe/Mg versus Si/Mg using Eq. 13 from \citet{unterborn2022nominal}, assuming that all the iron of the planet is in the core and that the mantle is composed of SiO$_2$ and MgO. This plot is similar to Figure 7 from \citet{unterborn2022nominal}, but here we plot only stars whose abundances [Fe/H], [Mg/H], [Si/H] are known with uncertainties smaller than 30\% and convert these values to molar ratios using the method by \citet{2022AJHinkel}. }
	\label{fig:star_ab}
\end{figure}

We assume that the high densities of high-density exoplanets are diagnostic of an anomalously high abundance of metals and do not indicate that these planets are compressed iron-rock-ice cores of gas giants (we revisit this alternative proposal in Section \ref{sec:conclusion}). Based on this assumption, we define a high-density exoplanet as a planet that has bulk density, $\rho$, higher than $\rho_{HD}$, which is the density corresponding to the 99.7th percentile of densities predicted for planets based on the abundance of iron and rock-forming elements measured in stars. Following \citet{unterborn2022nominal}, we define $\rho_{HD}$ (in cgs units) as

\begin{equation}
\label{eq:high-density_treshold}
    \rho_{HD} = 4.6 + 1.5~\bigg(\frac{R}{R_\oplus}\bigg)^{2.1} \sim 1.2~\bar{\rho},
\end{equation}

\noindent
where $\bar{\rho}$ indicates the density of a planet of the same radius, $R$ (in units of Earth radii, $R_\oplus$) and a bulk composition with the mean local galactic abundance of iron and rock-forming elements. The latter is close to that of CI carbonaceous chondrites (Figure \ref{fig:star_ab}). Planetary bodies with a bulk composition like CI chondrites are expected to differentiate into a rocky mantle and an iron core with a core-mass fraction, $Z$, of about 0.33, which is close to Earth's \citep{2003TrGePalme}. Differentiated iron-rock planets with $Z\sim$ 0.2 and $Z\sim$ 0.4 would instead have an iron content 1 standard deviation ($\sigma$) below and above the CI chondritic value, respectively (dashed lines tangent to the 68\% isocontour in Figure \ref{fig:star_ab}). 

To identify high-density exoplanets in the NASA Exoplanet catalog as of February 23, 2024, we looked for planets that orbit a single star and have radii and masses known with uncertainties smaller than 30\%, that is, $\sigma_R/R<$ 0.3 and $\sigma_M/M<$ 0.3 %
\citep[as in previous studies,][]{2021SciAdibekyan,unterborn2022nominal}. Next, we sampled 10$^5$ mass-radius pairs within the uncertainty ellipses of these planets, computed the corresponding $\rho$ assuming a spherical body and declared the planet ``high-density'' if the fraction $p_{HD}$ of mass-radius pairs for which $\rho > \rho_{HD}$ is larger than or equal to 68\% (that is, these planets are high-density at least at the 1 $\sigma$ confidence level). This search yielded a preliminary catalog of 13 high-density exoplanets.

Our preliminary catalog of high-density exoplanets includes three planets whose masses were estimated from analysis of transit-timing variations (TTVs): Kepler-245 c, Kepler-105 c, and Kepler-80 d. Previous studies have shown that analysis of TTVs for systems near first-order mean motion resonances may overestimate the planetary mass when the estimated eccentricities are high \citep[e.g., see Eqs. 6--7 in][]{2014ApJHaddenLithwick}. Kepler-245 c and Kepler-105 c were respectively flagged as high-eccentricity and low-eccentricity by \citet{2014ApJHaddenLithwick}. The mass of Kepler-80 d was derived by restricting the TTV fits to the low-eccentricity regime by \citet{2016AJMacDonald}, who run robustness tests to show that for the Kepler-80 system the mass–eccentricity degeneracy does not preclude reliable mass estimates. Following this, we decided to remove only Kepler-245 c from the aforementioned preliminary list. This shortened the list to 12 high-density exoplanets, whose $M$, $R$, $\rho$ and $p_{HD}$ are in Table \ref{tab:metal_rich_Pgi}.

The bulk densities of high-density exoplanets in Table \ref{tab:metal_rich_Pgi} range from 1.4 to 2.7 times those of similar-sized planets with an Earth-like composition. Under the assumption that these densities are proxies for high metal content, 
high-density exoplanets are differentiated in an iron-rich core overlaid by a thin rocky mantle made of rock-forming elements. This interior structure resembles that of planet Mercury \citep{2018mvambookEbel}, whose iron content is also above the 99.7th percentile of values expected from stellar cosmochemistry (Figure \ref{fig:star_ab}) with $Z\sim$ 70\% as measured by the NASA MESSENGER mission \citep[e.g.,][]{2018mvamNittler}. This has motivated the term ``super-Mercuries'' commonly used in the literature to refer to high-density exoplanets.

\subsubsection{Occurrence rate of high-density exoplanets}
\label{sec:observed_occurrence_rate}

The high ($\gtrsim$50 wt.\%) iron content expected for high-density exoplanets makes them the iron-rich end-members of the terrestrial planets, a family of astronomical bodies whose interior structure is thought to consist of an iron core overlaid by a rocky mantle, with a generally small (but hard to quantify, see below) amount of lower-density compounds (Figure \ref{fig:star_ab}). We aim to compute the occurrence rate of high-density exoplanets among the terrestrial planets, $\eta_{HD}$, and compare this metric to that from our simulations to test the hypothesis that mantle-stripping giant impacts between terrestrial planets occur at rates sufficient to explain the currently estimated abundance of the high-density exoplanets.

Estimating $\eta_{HD}$ requires identifying those planets in the catalog that are terrestrial based on their measured masses and radii. This is challenging for two reasons. First, there exists a significant degeneracy of compositions that can match a planet's measured mass and radius even when uncertainties are not taken into account. This prevents quantification of the exact amount of iron, rock and lower-density compounds in exoplanets. Second, how the relative abundances of iron, rock and lower-density compounds would affect the measured radii and masses depends on their chemical phase, miscibility and partition coefficients. Surface water could be present in the form of liquid or ice, or even be dissolved in the interior \citep{2024arXiv2Luo}. Furthermore, elements that are typically considered to be atmophiles (e.g., H, C) may partition into the core of terrestrial planets at the few wt.\% level, as it has been inferred for Earth and Mars \citep[e.g.,][]{mcdonough2003compositional,2023PNASIrving}.

Nevertheless, exoplanetary studies have long predicted that there should be a radius gap between ``volatile-rich'' planets and terrestrial planets, owing to atmospheric evaporation driven by the high-energy photons from stars, which is more efficient on smaller planets at a given stellocentric distance \cite[see][for a review]{owen2019atmospheric}. Consistently, exoplanetary observations have revealed a scarcity of planets with $1.5 R_\oplus \lesssim R \lesssim 2 R_\oplus$, known as the radius valley \citep[e.g.,][for a review; whether the radius valley is indeed due to photoevaporation or instead core-powered mass loss is still debated]{2021ARAAZhu}. Furthermore, the mass-radius distributions of exoplanets was found to change slope at the radius valley, suggesting a transition in the interior structure dominated by iron/rock to one dominated by volatiles \citep[e.g.,][and references therein]{2020AAOtegi}. We therefore leveraged the discovery of the radius valley to compute the value of $\eta_{HD}$ according to two criteria. In both cases, we used data from NASA Exoplanet Archive as of February 23, 2024 of planets that orbit a single star and have $\sigma_R/R<$0.3 and $\sigma_M/M<$0.3.

First, we assumed that the upper edge of the radius valley (that is, $R\approx 2R_\oplus$) is the maximum possible radius for a terrestrial planet. Based on this criterion, we identified a sample of 85 terrestrial planets which includes the 12 high-density exoplanets identified in Section \ref{sec:HD_exoplanets}, corresponding to $\eta_{HD} = 12/85 = 14\%$. Second, we defined a terrestrial planet as having $M$ and $R$ within ${\pm}1\sigma$ of those predicted by the mass-radius relationship for ``rocky planets'' by \citet{2020AAOtegi}. We used their mass-radius relationship to compute the radius given the measured mass, obtaining a sample of 84 terrestrial planets, to which we added the 12 high-density exoplanets whose radii for a given mass are too small to be captured by the mass-radius relationship by \citet{2020AAOtegi}. This yields $\eta_{HD} = 12/(84+12) = 12.5\%$.

The average $\eta_{HD}$ calculated from the criteria above is $\sim$13\%. We note that a previous study reported a higher $\eta_{HD}$ = 22\% based on the detection of 5 high-density exoplanets among a group of 22 terrestrial planets \citep{2021SciAdibekyan}. However, their sample included K2-38 b, for which we estimated a $p_{HD}$ of 55\%, lower than the $1\sigma$ threshold adopted here, and K2-106 b, whose nature as a high-density exoplanet has been recently revisited \citep{2023AJMartinez}. If we reject both K2-38 b and K2-106 b from the sample by \citet{2021SciAdibekyan}, their $\eta_{HD}$ lowers to 15\%, closer to our estimate.

Finally, we lowered $\eta_{HD}$ from 13\% to 9\% to account for the bias that the radial-velocity (RV) signal of higher-mass planets is stronger than that of less massive, similar-sized planets. This effect may bias the RV method towards providing better constraints of the mass of high-density exoplanets than for less massive exoplanets of the same size, assuming all other properties, e.g., orbit period, being equal \citep[e.g.,][]{2016MNRASSteffen}. We describe the procedure that we followed to correct $\eta_{HD}$ for the RV bias in Appendix \ref{appx:RV_bias} and Figure \ref{fig:RV_bias}.

\subsection{Model of orbital instability of compact systems}
\label{subsec:statistical model}

\subsubsection{\textit{N}-body simulations}

The orbital dynamics and collisional evolution of planetary systems is traditionally studied using \textit{N}-body simulations \citep[e.g.,][]{2015ApJVolkGladman,2020PNASTamayo}. These simulations evolve the orbits of planets under their mutual gravitational interaction and model close encounters and collisions. To showcase this, we simulated the orbital evolution of the four super-Earths in the compact system Kepler-402 using \texttt{REBOUNDx} including post-Newtonian general relativity orbital correction using the \texttt{gr}-package. We initialized the eccentricity and mean longitudes of these planets with values that are predicted to lead to instability according to the Stability of Planetary Orbital Configurations Klassifier, SPOCK \citep{2020PNASTamayo}. Initial eccentricities were kept below 0.1, consistent with constraints from transit timing variations \citep{2017AJHadden}. Any collision between the planets was modelled using the collision model described in Section \ref{sec:collision_model}.

\begin{figure}[p]
    \centering
\includegraphics[width=\linewidth]{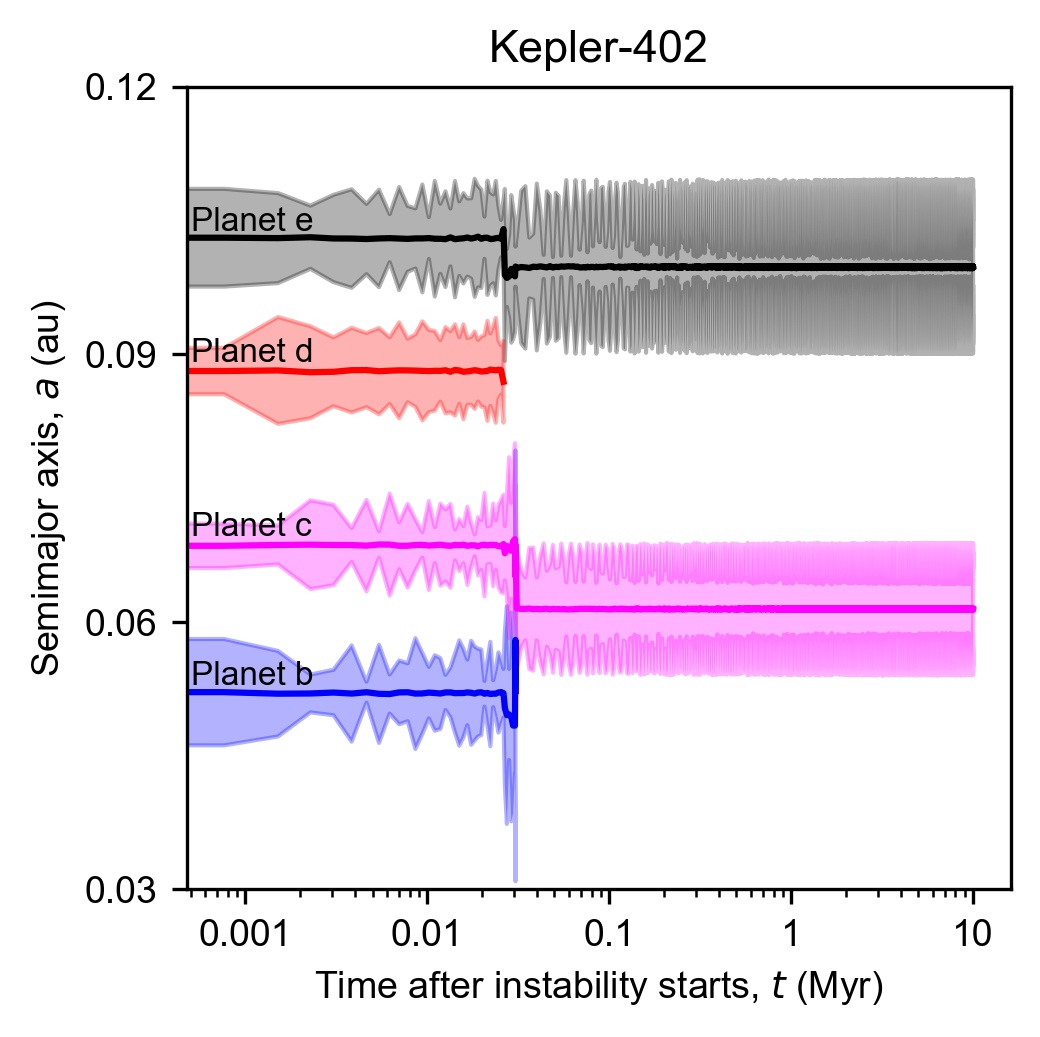}
    \caption{\textbf{Instability of a compact planetary system.} \textit{N}-body evolution of the semi-major axis, $a$, of the super-Earths in the compact system Kepler 402 as a function of time, $t$. The solid curves indicate the semimajor axes and the envelopes are the periapses and apoapses of the orbits. Upon instability, planet b merges with planet c and planet d merges with planet e by giant impacts. No planet becomes metal rich in this simulation, but a different outcome is possible for different initial eccentricities and mean longitudes.}
	\label{fig:N-body}
\end{figure}

The result for Kepler 402 is in Figure \ref{fig:N-body}, which plots the evolution of the semimajor axis, $a$ [measured in astronomical units (au)] as a function of time. The orbital instability features the two innermost planets and the two outermost planets that merge pairwise to form two larger planets. No planet becomes metal-rich in this simulation (that is, no planet acquires a bulk density higher than $\rho_{HD}$, defined in Eq. \ref{eq:high-density_treshold}), but a different outcome is possible for different initial eccentricities and mean longitudes. However, using \textit{N}-body simulations to comprehensively explore the parameter space of initial orbits across multiple compact systems and effectively couple \textit{N}-body and SPH simulations is currently computationally impossible \citep{2020PNASTamayo,2023AREPSGabrielCambioni}. To overcome this challenge, we developed and run a statistical approach.

\subsubsection{Statistical model}
\label{sec:statistical_model}

\begin{figure}[p]
	\centering
	\includegraphics[width=1\linewidth]{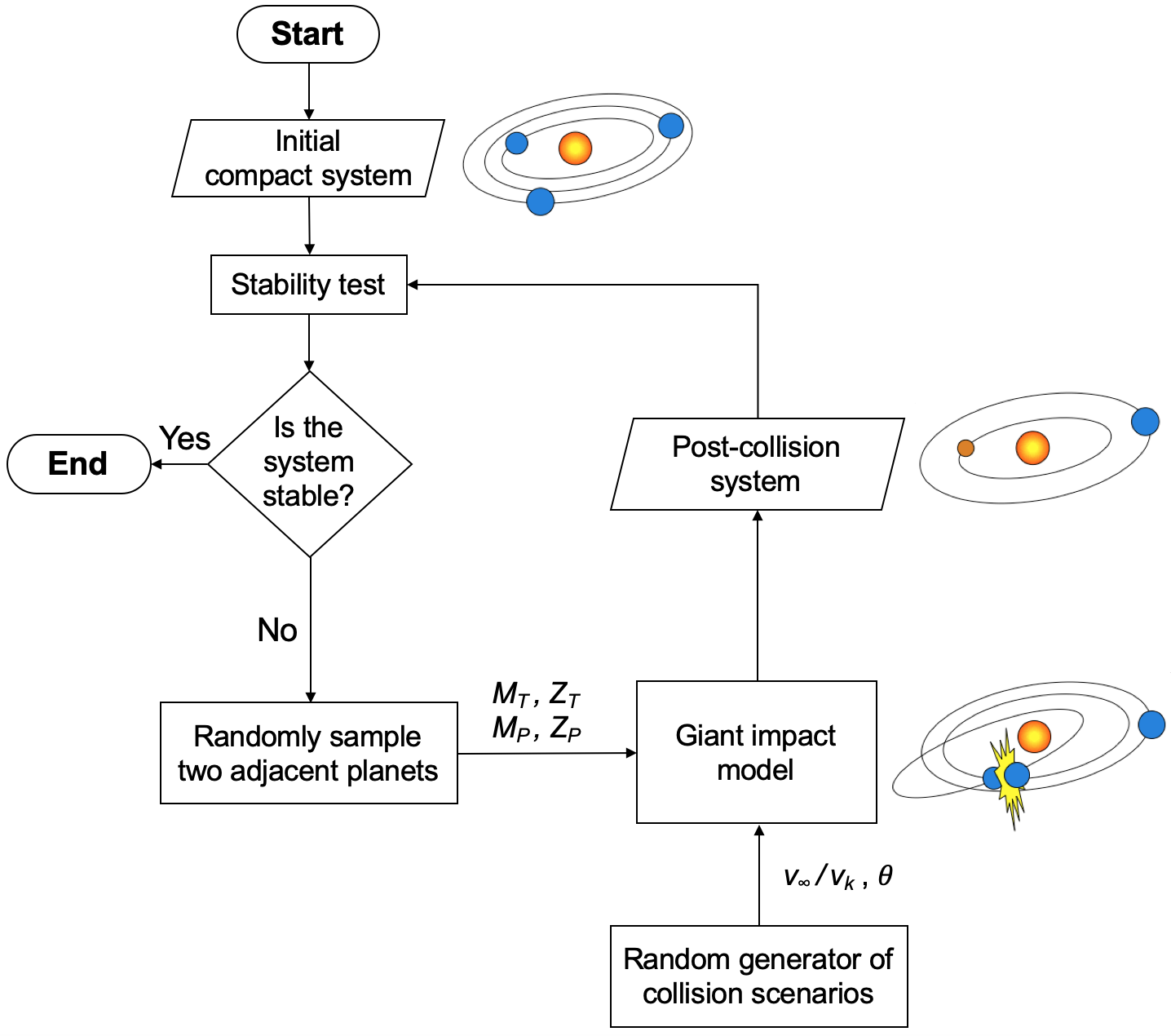}
	\caption{\textbf{Flowchart of the statistical model of orbital instability}. For details on each step, see Section \ref{sec:statistical_model}. The symbols $M_T$ and $M_P$ indicate the masses of the target and projectile in the collisions, respectively. The symbols $Z_T$ and $Z_P$ indicate the core-mass fractions of the target and projectile of the collisions, respectively. The distributions from which the relative impact velocity, $v_\infty$ (in units of Keplerian velocity, $v_k$) and impact angle, $\theta$, are sampled are shown in Figure \ref{fig:input_conditions}c,d. }
	\label{fig:flowchart}
\end{figure}

Our statistical model of orbital and collisional evolution of compact systems is composed of three steps (Figure \ref{fig:flowchart}).

First, we determine the stability of a compact system as a function of orbital eccentricities, $e$, randomly sampled from an uniform distribution between 0 and 0.1 \citep[as constrained by the analysis of transit timing variations, ][]{2017AJHadden}, longitudes of pericenter, $\varpi$, and true anomalies randomized between 0 and 2$\pi$ using well-established analytical and statistical tools, assuming coplanar orbits. We declare a system of two planets unstable if the difference in their complex orbital eccentricities, $\Theta$,

\begin{equation}
\Theta = \frac{1}{\sqrt{2}}~[e_2 \exp{(j\varpi_2)}-e_1 \exp{(j\varpi_1)}],
\end{equation}

\noindent
is higher than the instability threshold, $\Theta^{*}$ \citep{2018AJHadden}:

\begin{equation}
\label{eq:inst_criterion}
\Theta^{*} = \frac{a_2-a_1}{a_1} \exp{\bigg[-2.2(\mu_1+\mu_2)^{1/3}\bigg(\frac{a_2}{a_2-a_1}\bigg)^{4/3}~\bigg]},
\end{equation}

\noindent
where 1 and 2 indicate the inner and outer planets respectively and $\mu$ is planet-to-star mass ratio. If the system has 3+ planets, we reject the null hypothesis that the system is stable if the probability of stability computed using the SPOCK algorithm by \citet{2020PNASTamayo} is smaller than 0.05. We note that SPOCK was trained to classify the stability of systems over 10$^9$ orbits of their innermost planet (typical orbital period $<$ 10 days), which is shorter than the typical age of the host stars of $\sim$1--10 Gyr \citep{2017AJJohnson}. The fraction of compact systems expected to experience instability during every order of magnitude of time is predicted to be roughly constant and $<$7\% \citep{2015ApJVolkGladman,2020AJVolk,2021AJYee}. As such, by using SPOCK, we may underestimate the fraction of unstable systems by $<$14--20\% because the system ages are two to three orders of magnitude older than the timescale for SPOCK instabilities.

Second, for a given unstable system, we postulate that the two adjacent planets with the smallest separation in units of mutual Hill radii undergo a collision. The value of $\theta$ is randomly sampled from probability distributions derived from geometric arguments and the value of $v_{\infty}$ is computed as 

\begin{equation}
\label{eq:chi_value}
    v_{\infty}=\chi~v_K,
\end{equation}

\noindent
where $\chi$ is a dynamical excitation factor randomly sampled from probability distributions derived from \textit{N}-body simulations of orbital instability (details on the $\theta$ and $\chi$ distributions are provided in Section \ref{sec:initial_conds}). 
$v_K$ is the Keplerian velocity computed at the stellocentric distance of the collision, $a_{coll}$. The value of $a_{coll}$ is randomly sampled from a uniform distribution in the range [$a_T$, $a_P$]. To realistically model the collision with a runtime of seconds, we developed a machine-learning algorithm that mimics the outcome of high-resolution, computationally expensive SPH simulations of giant impacts, building upon our previous studies \citep{2019ApJCambioni,2020ApJEmsenhuberA,2021PSJCambioni}. This collision model is described in Section \ref{sec:collision_model}.

Third, after the collision we update the masses, core-mass fractions and stellocentric orbits of the largest remnant, any second-largest remnant and any debris particles based on the output of the collision model and Eqs. 2.134-2.140 from \cite{2000BookMurrayDermott}. If the total orbital energy of the system after the collision, $E$, is greater than the pre-impact value, $E_0$, the model resets the system to its pre-collision state and runs a new collision until $(E-E_0)$/$E_0\geq$ 0 is obtained with a tolerance of 1\%. This tolerance is to account for error propagation from the giant-impact model predictions, while enforcing the energy criterion to at least 1 part in 10, as suggested by previous studies \citep{2015ComACBoekholt}. Next, those planetary bodies that are either ejected from the system or accreted by the central star ($a<a_{Roche}$, where $a_{Roche}$ is the Roche's limit of the host star/planet pair) are removed from the system. Finally, the model assesses the stability of the post-collision system using Eq. \ref{eq:inst_criterion} or SPOCK, depending on the system multiplicity. The collision step is repeated until the system becomes stable or achieves multiplicity equal to 1. We reject simulations for which, overall, $(E-E_0)$/$E_0<$ -5\%.

\subsubsection{Collision model} 
\label{sec:collision_model}

To model giant impacts in our  simulations, we designed seven neural networks, $g_i$ ($i$ = 1,...,7), that approximate giant impacts outcomes $y_i$ ($i$ = 1,...,7), as a function of the array of impact conditions, $x$:

\begin{equation}
    \begin{cases}
      y_1 = (\xi_L, \xi_S) &= g_1 (x)~~~~\forall~\text{collisions,}\\
      y_2 = Z_L &= g_2 (x)~~~~\forall~\text{collisions,}\\
      y_3 = (\beta, \xi_{d}) &= g_3 (x)~~~~ \forall~\text{collisions,} \\
      y_4 = Z_S &= g_4 (x)~~~~ \forall~\text{hit-and-run collisions,}\\
      y_5 = \epsilon' &= g_5 (x)~~~~ \forall~\text{hit-and-run collisions,} \\
      y_6 = b' &= g_6 (x)~~~~ \forall~\text{hit-and-run collisions,}\\
      y_7 = \Delta \omega' &= g_7 (x)~~~~ \forall~\text{hit-and-run collisions,}
    \end{cases}
\end{equation}

\noindent
where the elements of $x$ are: logarithm of the target's mass, $M_T$; logarithm of the projectile's mass, $M_P$; core-mass fractions of target and projectile, $Z_T$ and $Z_P$, respectively; $v_{coll}/v_{esc}$; and the sine of the impact angle, $\theta$. The impact outcomes $y_i$ are: the accretion efficiencies of the collision, $\xi_L$ = ($M_L-M_T)/M_P$ and $\xi_S$ = ($M_S -M_P)/M_P$, where $M_L$ and $M_S$ are the masses of the largest remnant and any escaping projectile in hit-and-run collisions \citep[termed ``second-largest remnant'' hereafter; we declare a collision to be a hit-and-run if $M_S<0.1M_P$ as in the previous study by][]{2020ApJEmsenhuberA}; the core-mass fractions of the largest remnant and any second-largest remnant, $Z_L$ and $Z_S$, respectively; the specific energy, $\epsilon'$, post-collision impact parameter, $b'$, and variation of the argument of pericenter, $\Delta \omega'$, of the egress hyperbolic relative orbit of any second-largest remnant, as in \cite{2020ApJEmsenhuberA}; the power-law index, $\beta$, of the mass-frequency distribution for debris smaller than any largest remnants and the mass fraction of largest debris body, $\xi_{d}$, computed over a total debris mass $M_D$ = $M_T+ M_P- M_L- M_S$.

The networks were trained, validated and tested using a look-up table of impact conditions and outcomes built by post-processing a state-of-the-art database of 1250 high-resolution SPH simulations of giant impacts from \cite{2023Emsenhuber_inrev}. The colliding bodies in the dataset are planets differentiated into a dunite mantle and an iron core whose $Z$ varies between 0 and 1. The target's mass spans 6 orders of magnitude, from 10$^{-4}$ to 5 $M_\oplus$. We used 70\% of the table entries for training using the \texttt{Adam} algorithm \citep{kingma2014adam}. Another 15\% of the table entries were used to identify the network architectures that do not overfit the training data by combining the following hyperparameters: learning rates of 0.002, 0.02, or 0.2; up to three hidden layers with 16, 32, 64, 128, or 256 neurons per layer; ReLu, Sigmoid, or Tanh activation functions; drop-out rates of 0.2\%, 2\%, or 20\%; input normalization following the Standard, MinMax, or Robust scalings. To minimize overfitting, we operated Batch Normalization on the input layer and added Gaussian noise with strength 0.005 prior to the output layer. We used the remaining 15\% of the table entries to assess the networks' performance on unseen data in units of mean squared error (MSE) between the networks' predictions and the SPH predictions.

Using the procedure described above, we generated 66295 converged neural networks, from which we chose the optimal networks $g_i$ ($i$ = 1,...,7)  as those that do not overfit the training data and have the lowest MSE at testing (Figure \ref{fig:NNPerformance}).  
These networks generalise well to the prediction of the giant-impact outcomes at testing, with correlation coefficients $>$89\%. Furthermore, the networks $g_1$ and $g_2$ accurately extrapolate $\xi_L$ and $Z_L$ for $M_T>$ 5 $M_\oplus$ (that is, the upper limit of the training database) with MSEs equal to 0.01 and 0.02, respectively, which are comparable to the SPH numerical noise level \citep{2023Emsenhuber_inrev}. Here the MSEs were computed between $\xi_L$ and $Z_L$ from networks $g_1$ and $g_2$, and $\xi_L$ and $Z_L$ predicted by \citet{2022MNRASReinhardt}, who modelled collisions using a different SPH code than \citet{2023Emsenhuber_inrev}. For this comparison, we assumed $M_T$ between 1 and 20 $M_\oplus$, $\theta$ = 0$^\circ$, $Z_T = Z_P = $ 0.3, and $M_T = M_P$ and explored impact energies between 0 and 2 times the energy for catastrophic disruption computed using Eqs. A.1 and A.3 from \citet{2016IcarusMovshovitz}.

We implemented the neural networks in a collision resolver which improves upon our previous work \citep{2019ApJCambioni,2020ApJEmsenhuberA,2021PSJCambioni} by including explicit production and orbital evolution of debris smaller than the second largest remnant and self-consistent evolution of planetary core-mass fractions in impact chains. Based on $M_L$, $M_S$, $Z_L$ and $Z_S$, the collision resolver predicts the radius of the impact remnants using a interpolating function based on the masses, radii and core-mass fractions of the initial SPH bodies in the database by \citet{2023Emsenhuber_inrev}. For the debris, the collision resolver computes $M_D$ and the metal content of debris $Z_D$ by imposing conservation of mass. If $Z_D > 1$ due to accumulation of numerical errors, then $Z_D$ is randomly sampled from a Rayleigh distribution with scale value of 0.1 to account for the prediction by \citet{2021PSJCambioni} that $\sim$ 85\% of giant-impact debris are composed of mantle materials; $Z_L$ and $Z_S$ are updated by imposing conservation of mass. If $v_{coll} \gtrsim$ 2 $v_{esc}$, debris tends to be vaporized by the hydrostatic pressure release \citep{2014NatGeoAsphaug,2023Emsenhuber_inrev}, such that in this case we assume that the largest remnant accretes the whole debris mass. Otherwise, giant impacts tend to produce sizable debris fragments \citep{2023Emsenhuber_inrev}, and we add two debris bodies with masses $M_{d1}>M_{d2}>$ 10$^{-4}$ $M_\oplus$ and metal fractions $Z_{d1} = M_{d1}/M_D~Z_D$ and $Z_{d2}=M_{d2}/M_D~Z_D$ to the post-collision planetary population and assume that the remaining debris mass is accreted by the largest remnant. Masses $M_{d1}$ and $M_{d2}$ are randomly sampled from the mass-frequency distribution predicted by the neural network $g_3(x)$. The clumps are ejected radially from the edge of the largest remnant's Hill sphere with $v_{\infty}$ = 0.5 $v_{coll}$ following \citet{2023Emsenhuber_inrev}. Giant impacts with $M_P < 0.05 M_T$ or $M_T<$ 10$^{-4}~M_\oplus$ are modelled as perfectly inelastic.

\subsection{Compact systems: observations and orbital dynamics}
\label{sec:initial_conds}

To initialize the model of orbital instability of compact systems described in Section \ref{sec:statistical_model}, we built a dataset of observed compact systems by searching in the NASA Exoplanet Archive for planetary systems containing multiple planets with orbital periods between one day and one year. We limited our search to planets around single stars and that have a terrestrial composition (defined as having $R_0<$ 2$R_\oplus$, see Section \ref{sec:observed_occurrence_rate}; hereafter, we use the subscript ``0'' to indicate the properties of the super-Earths in the compact systems before the instability.) This assumption is supported by the observation that 70\% of planets in the systems with high-density exoplanets are terrestrial planets (Figure \ref{fig:HDsystems} and Section \ref{sec:formation_mechanisms}).

If the mass of a super-Earth in these compact systems, $M_0$, is unknown or known with $\sigma_{M_0}/M_0>$ 30\%, we set the planet's initial core-mass fraction, $Z_0$, to be 0.33. This value corresponds to a bulk composition with the average local galactic abundance of iron and rock-forming elements (Figure \ref{fig:star_ab}). Otherwise, we derived $Z_0$ from the observed radius, $R_0$, and $M_0$ using the interpolation function based on the pre-impact bodies of the SPH simulations used to train the collision model of Section \ref{sec:collision_model}. We rejected those systems in which at least one planet is a water-rich world or a high-density world \citep[that is, if there are planets with density at least 3 standard deviations below or above the expected range of galactic values, ][]{unterborn2022nominal}. 
The cumulative distributions of the masses and radii of the super-Earths in compact systems are plotted in Figure \ref{fig:input_conditions}a,b. 

\subsubsection{Model scenarios}
\label{subsec:model_variations}

In our nominal scenario N0, we assumed that the precursor systems of the observed high-density exoplanets are well represented by the compact systems in our dataset (we provide evidence supporting this assumption in Section \ref{sec:sys_instability}). For the collision model, we randomly sampled the value of $\chi$ (which defines $v_{\infty}$ and thus $v_{coll}$ given $v_K$ and $v_{esc}$, Eqs. \ref{eq:v_coll} and \ref{eq:chi_value}) from its probability distribution derived from previously published results of \textit{N}-body simulations by \citet{2015ApJVolkGladman}, who evolved the orbits of selected compact systems until they experienced collisions. In addition, we randomly sampled the impact angle $\theta$ from the differential probability $dP(\theta)$ = $sin(2\theta)$ $d\theta$, which yields a most likely $\theta$ of 45$^\circ$. This probability function was first derived by \citet{1893mfsobookGilbert} from the geometric argument that the differential probability that a point-mass impactor would hit a massless spherical target at a distance $x=R_T\sin{\theta}$ from its center is $dP(\theta)$ = $2\pi~x~dx$. \citet{1962BookShoemaker} later showed that the proportionality $dP(\theta)\propto \sin{2\theta}$ holds when the gravity of the largest body is included. This distribution is expected to hold for giant impacts too \citep{2010ChEGAsphaug}. We refer to \citet{2023AREPSGabrielCambioni} for a full derivation of $dP(\theta)$.

Because the properties and dynamical environment of the precursor planetary systems are a priori unknown, we explored variations of the nominal scenario aimed at encompassing a large diversity of scenarios and corresponding model parameters. Each scenario described below is identical to the nominal scenario except for one parameter.

\begin{enumerate}

\item In scenario T1, we excluded from the catalog of precursor systems those systems that have two initial planets, which \citet{2015ApJPuWu} and \citet{2015ApJVolkGladman} suggested might be the descendants of more closely packed planetary systems that already experienced instability \citep{2015ApJPuWu,2015ApJVolkGladman};

\medskip

\item In scenario T2, we assumed that the super-Earths in the precursor systems whose mass is unknown or poorly constrained ($\sigma_{M_0}/M_0>30\%$) are metal-poorer than the nominal scenario. As such, we initialized the masses of these super-Earths assuming $Z_0 = 0.2$, which corresponds to an iron content 1 standard deviation below the average galactic value (Figure \ref{fig:star_ab});

\medskip

\item In scenario T3, we assumed that the super-Earths in the precursor systems whose mass is unknown or poorly constrained ($\sigma_{M_0}/M_0 > 30\%$) are more metal-rich than the nominal scenario. As such, we initialized the masses of these super-Earths assuming $Z_0 = 0.4$, which corresponds to an iron content 1 standard deviation above the average galactic value (Figure \ref{fig:star_ab});

\medskip

\item In scenario T4, we assumed that the probability distribution of $\chi = v_\infty/v_k$ follows a Rayleigh distribution with a scale value of 0.1, corresponding to a lower degree of dynamical excitation than in the nominal scenario (Figure \ref{fig:input_conditions}c);

\medskip

\item In scenario T5, we assumed that the probability distribution of $\chi = v_\infty/v_k$ follows a Rayleigh distribution with a scale value of 0.4, corresponding to a higher degree of dynamical excitation that in the nominal scenario (Figure \ref{fig:input_conditions}c);

\medskip

\item In scenario T6, we relaxed our assumption that every super-Earth in the precursor systems must have $R_0 < 2 R_\oplus$ and allowed planets with $2 R_\oplus < R_0 < 3 R_\oplus$ into the sample. This scenario is motivated by the recent detection of planets with radii between 2 and 3 $R_\oplus$ and densities consistent with a volatile-poor composition \citep{2020AAOtegi}. Based on this, we found 12 additional precursor systems and explored their stability and collisional evolution using our statistical model. 
\end{enumerate}

\subsection{Comparison with observed high-density exoplanets}
\label{sec:exosurvey}

We compared the simulated metal-rich worlds with the observed high-density exoplanets using three metrics: (1) predicted versus observed occurrence rate; (2) similarity between the modelled versus observed mass-radius distributions; and (3) the rate at which giant impacts can form metal-rich worlds as massive and large as the observed high-density exoplanets.

\subsubsection{Predicted occurrence rate}
\label{sec:eta_MR}

We computed the occurrence rate of simulated metal-rich planets formed in our model of orbital instability, $\eta_{MR}$, as their fraction in the post-instability planetary population. We correct $\eta_{MR}$ for typical biases in the exoplanetary population as:

\begin{equation}
\label{eq:predicted_MR}
    \tilde{\eta}_{MR} = \frac{\sum_{i=1}^n p_{survey}}{\sum_{i=1}^{n_{tot}} p_{survey}},
\end{equation}

\noindent 
where $n$ is the number of simulated metal-rich planets, $n_{tot}$ is the total number of post-instability planets, and $p_{survey}$ is the probability that transiting planets with $R$ and orbital period as the simulated metal-rich planets are reliable planet candidates. We calculate $p_{survey}$ using the Kepler Exoplanet Population Observation Simulator \citep{2019ApJMulders} for stars more massive than 0.6 Sun's masses and the Transiting Exoplanet Survey Satellite likelihood model \citep{2019AJBallard} otherwise. We estimated the predicted number of metal-rich planets in the terrestrial planet population as the number of terrestrial planets in our dataset, times $\tilde{\eta}_{MR}$. 

\subsubsection{Similarity of mass-radius distributions}
\label{sec:KS_test}

We tested the null hypothesis that the population of observed high-density exoplanets as a whole formed by mantle-stripping giant impacts by means the 2-dimensional Kolmogorov–Smirnov (KS) test (package: \url{github.com/syrte/ndtest}). This test assesses whether the mass-radius distributions of the simulated metal-rich worlds and that of the observed high-density exoplanets are drawn from the same underlying distribution. For this test, we built the mass-radius distribution for the simulated metal-rich planets by randomly sampling the ($M$, $R$) values of 2/3 of metal-rich planets from the model population with weights equal to $p_{survey}$, without replacement. We estimated the mass-radius distribution of the observed high-density exoplanets by bootstrapping 1,000 ($M$, $R$) values for each planet within uncertainties.

\subsubsection{Rate of formation of metal-rich worlds akin to the observed high-density exoplanets}
\label{sec:pgi}

The KS analysis of Section \ref{sec:KS_test} is meant to test the null hypothesis that mantle stripping by giant impacts is the predominant pathway through which the population of high-density exoplanets form. However, even in case of the KS test rejects this null hypothesis, it is still possible that some of the observed high-density exoplanets formed through mantle-stripping giant-impacts. To investigate this, we computed the rate $p_{GI}$ at which a given scenario among those described in Section \ref{sec:initial_conds} form simulated metal-rich worlds with $M$ and $R$ similar to each of the observed high-density exoplanets. We compute $p_{GI}$ as the fraction of simulated metal-rich worlds with $M$ and $R$ within the 3-standard-deviation uncertainty ellipse of the $M$ and $R$ of the observed high-density exoplanets. 

\section{Results and Discussion}
\label{sec:results}
\subsection{Instability of compact systems}
\label{sec:sys_instability}

\begin{figure}[p]
	\centering
	\includegraphics[width=0.95\linewidth]{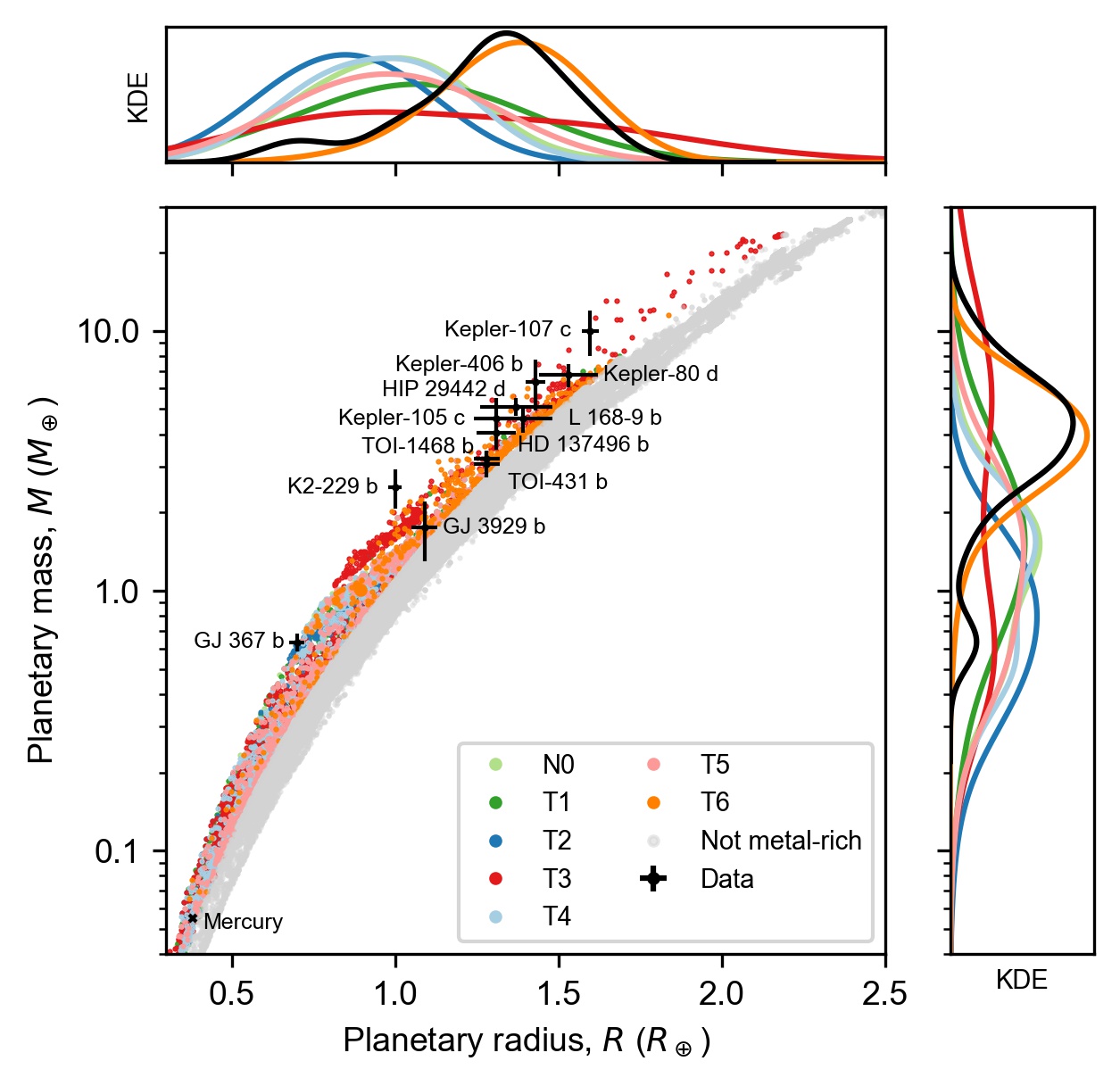}
	\caption{\textbf{Results of the statistical model for different model assumptions}. Mass, $M$, as a function of radius, $R$, of the simulated planets surviving an orbital instability for the different model assumptions described in Section \ref{sec:initial_conds}. The final planets are deemed metal-rich remnants (colored symbols) and non--metal-rich remnants (grey symbols) based on the high-density threshold of Eq. \ref{eq:high-density_treshold}. The black data points correspond to the observed high-density exoplanets listed in Table \ref{tab:metal_rich_Pgi}, with 1 standard deviation uncertainty bars; planet Mercury is plotted for comparison. The side plots are the Kernel Density Estimates (KDEs) of the radius (top) and mass (left) of the observed high-density planets (black curve) and the simulated metal-rich giant-impact remnants (colored curves). The model KDEs are weighted according to the probability of detection of the simulated metal-rich worlds by transit surveys, to mimic observational bias typical of exoplanetary surveys (Section \ref{sec:exosurvey}). For a version of this figure plotting the occurrence rates of the observed high-density exoplanets versus those of simulated metal-rich worlds of similar masses and sizes, see Figure \ref{fig:all_MR}.}
	\label{fig:all_MR_KDE}
\end{figure}

\begin{figure*}
	\centering
	\includegraphics[width=1\textwidth]{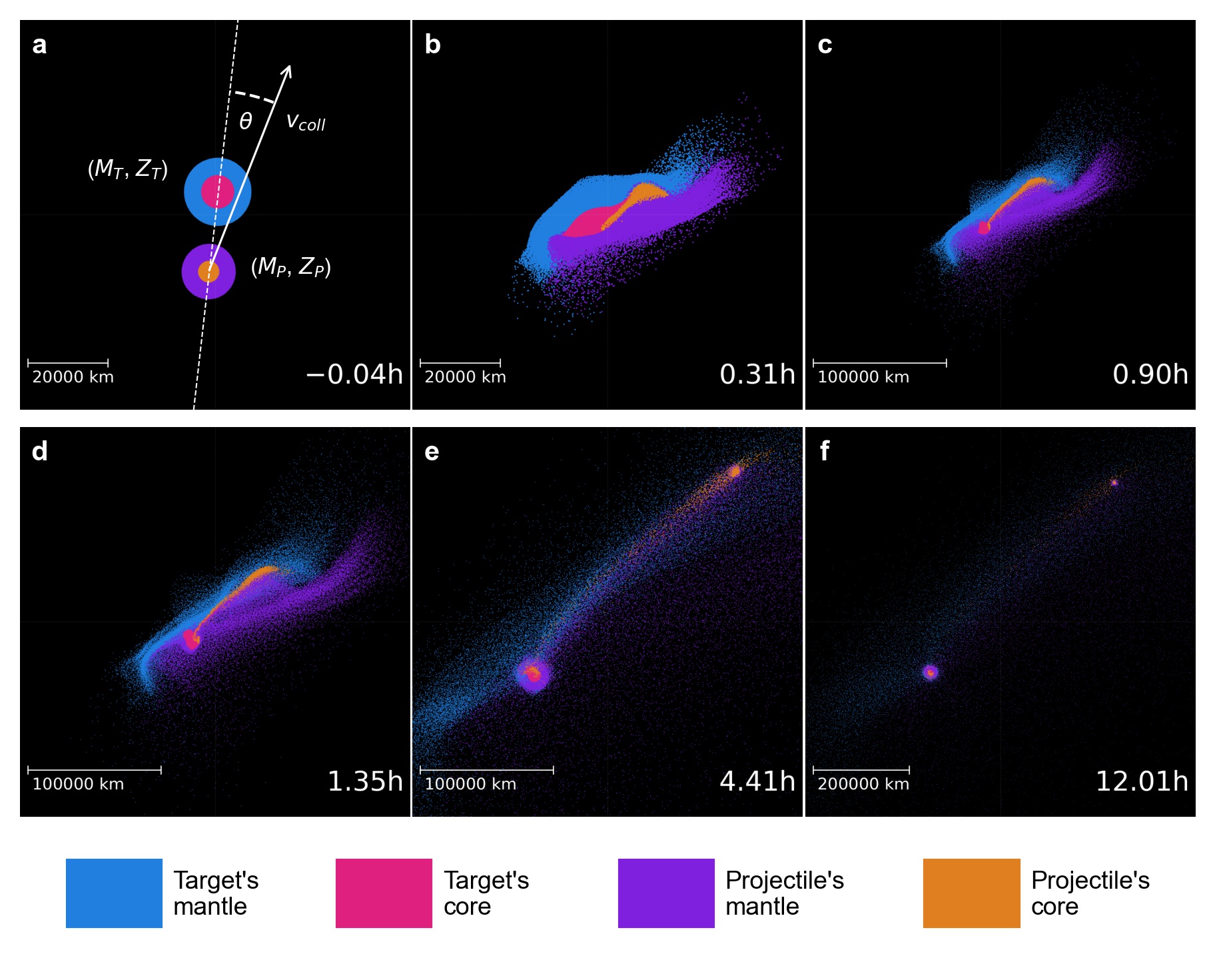}
	\caption{\textbf{A hit-and-run collision between two super-Earths forming a metal-rich planet}. \textbf{a--f}, snapshots of a SPH simulation of a giant impact (times after collision listed at lower right each panel) from the dataset by \citet{2023Emsenhuber_inrev}. The target body has mass $M_T$ = 2.5 $M_\oplus$ and core-mass fraction $Z_T$ = 0.30; particles composing the mantle and core of the target are colored in blue and red, respectively. The projectile has mass $M_P$ = 1 $M_\oplus$ and core-mass fraction $Z_P$ = 0.16; particles composing the mantle and core of the projectile are colored in purple and orange, respectively. The collision velocity is $v_{coll}$ = 2.54 $v_{esc}$ and the impact angle $\theta$ = 15$^\circ$ [see definitions in panel (a)]. Note scale bars at lower left. The projectile's mantle is accreted by the target, while the projectile's core escapes as a metal-rich planet.}
	\label{fig:SPH}
\end{figure*}

We find that 29$\pm$17\% of the precursor compact systems may become unstable within 10--100 Myr and thus experience orbit crossing and possibly a phase of giant impacts. In the nominal scenario N0, $\sim$20\% of the stable compact systems show mean-motion-resonances between planetary pairs (Figure \ref{fig:MMRs}), consistent with the general paucity of resonances and near-resonance features observed across compact systems \citep[][for a review]{2022PPVIIWeiss}. Importantly, in the nebular phase of planet formation, planet-disk interactions can lead planets to experience orbital migration, which if convergent may naturally shepherd planets into mean-motion resonances \citep[e.g.,][]{2015MNRASBatygin}. As such, the lack of resonances in the stable compact systems in our dataset suggests that these systems are stable today either because they escaped resonance during the nebular period or experienced an orbital instability after nebular gas dissipation  
\citep{2014AJGoldreichSchlichting,2017MNRASIzidoro,2021AAIzidoro,2020ApJMatsumotoOgihara}. 

By contrast, we find that 70\% of the compact systems that we predict to become unstable have at least a pair of super-Earths in near--mean-motion resonances (Figure \ref{fig:MMRs}). This suggests that at least 70\% of the compact systems in our catalog whose instability may form metal-rich planets are likely to be primordial (that is, they did not experience an instability yet). This builds confidence in our assumption that the majority of the unstable compact systems are representative of the possible precursor systems of the observed high-density exoplanets.

\subsection{Giant-impact outcomes}
\label{sec:formation_mechanisms}

Across all scenarios in Section \ref{sec:initial_conds}, we record a total of 109 unstable system configurations (Table \ref{tab:results_compact}). For each unstable configuration, we used the statistical model described in Section \ref{subsec:statistical model} to simulate its orbital and collisional evolution 1000 different times, for a total of 109$\times$1000 = 109'000 different simulations.

In most of these collision scenarios, the super-Earths in the compact systems merge to form larger planets without a significant change in metal content, potentially explaining the detection of planets with an Earth-like interior structure above the radius-gap valley \citep{2023AJEssack,2023NaponielloNature}. In some other cases, the super-Earths experience chains of $\sim$ 2--3 collisions that strip their mantles, causing them to evolve into metal-rich, high-density planets (Figures \ref{fig:all_MR_KDE} and \ref{fig:all_MR}). The giant impacts that form the simulated metal-rich planets are hit-and-run collisions in 99\% of the cases. In these events, a projectile super-Earth grazes a larger target super-Earth and continues along a deflected stellocentric orbit as a metal-rich planet distinct from the target (Figure \ref{fig:SPH}). The escaping projectile avoids reaccreting its own debris because its mantle is mostly accreted by the target following the collision. This scenario is analogous to the hit-and-run theory proposed to explain the anomalously large metallic core of planet Mercury \citep{2014NatGeoAsphaug}, with the difference that, in our model, the giant impacts involve super-Earths and occur at smaller stellocentric distances.

In our simulations, we assumed that collisions occur between super-Earths with no gaseous envelopes (Section \ref{sec:initial_conds}). However, in compact planetary systems, super-Earths may have sub-Neptune companions too \citep{2022PPVIIWeiss}. Indeed, five high-density exoplanets (TOI-1468 b, TOI-431 b, HIP-29442 d, Kepler-107 c and Kepler-80 d) have transiting sub-Neptune companions (Figure \ref{fig:HDsystems}). 

\begin{figure}[p]
	\centering
	\includegraphics[width=0.65\linewidth]{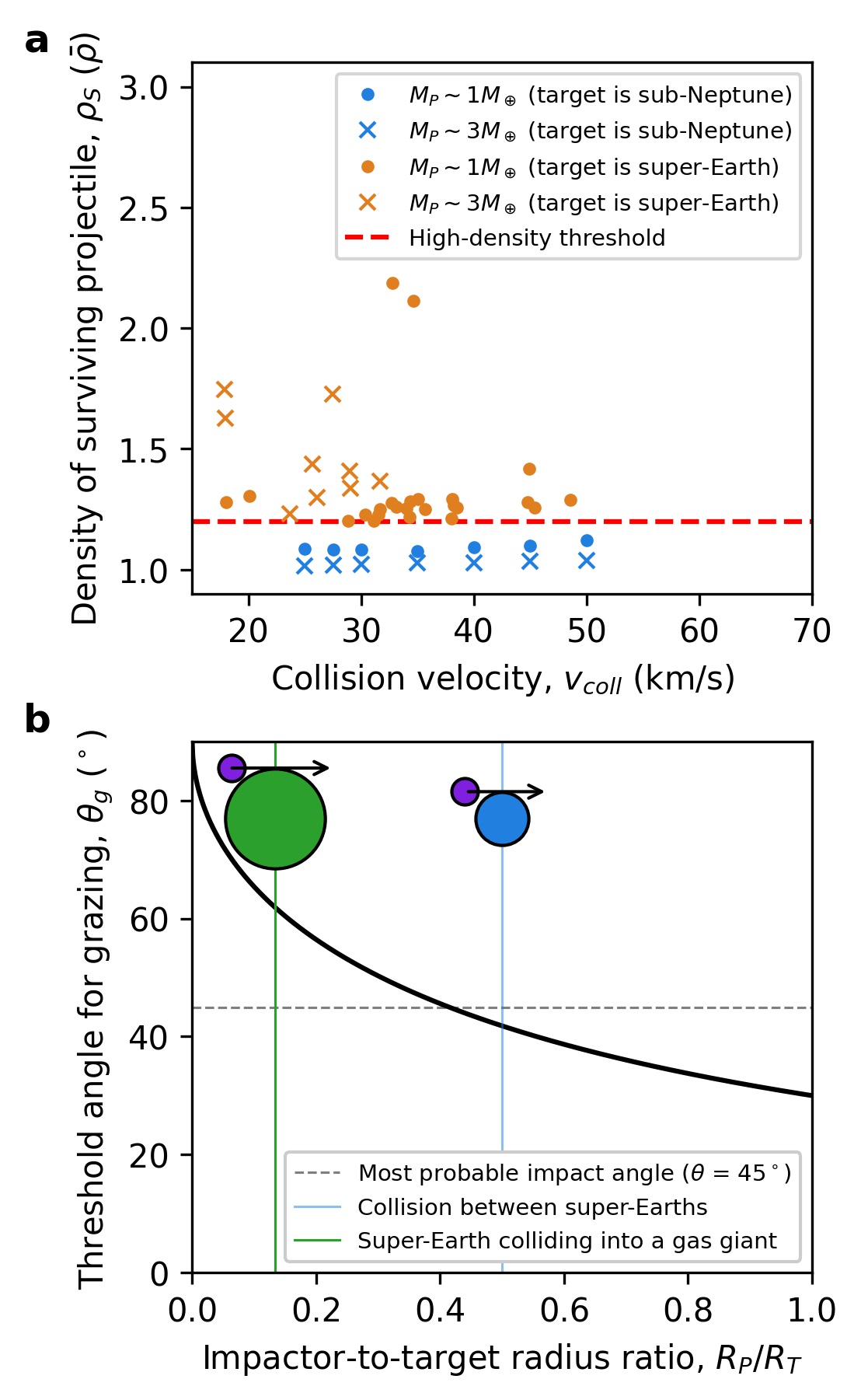}
	\caption{\textbf{Grazing collisions involving atmosphere-rich planets}. \textbf{a}, Density of the surviving hit-and-run projectile, $\rho_S$, as a function of $v_{coll}$. The value of $\rho_S$ is in units of $\bar{\rho}$, which is the density of a planet with the same $R$ and a bulk composition with the local galactic abundance of iron and rock-forming elements. We assume $\theta=$ 45$^\circ$. The predictions for the Sub-Neptune target are from the database of SPH simulations by \citet{2022MNRASDenman}. The predictions for the super-Earth target are from our collision model described in Section \ref{sec:collision_model}. \textbf{b}, Minimum impact angle, $\theta_g$, required for the geometric center of a projectile to graze the surface of its target \cite[Eq. 9 in][]{2010ChEGAsphaug} as a function of their projectile-to-target radius ratio, $R_P/R_T$. A super-Earth with $R_P$ = 1.5 $R_\oplus$ grazes a Jupiter-sized planet ($R_T$ = 11.2 $R_\oplus$) for $\theta>\theta_g$ = 60$^\circ$. By contrast, if the target is a super-Earth with $R_T$ = 2 $R_P$, $\theta_g$ is within 5$^\circ$ of the most probable impact angle of 45$^\circ$.}
	\label{fig:volatile-rich}
\end{figure}

Motivated by this, we assess the robustness of our assumption that all colliding bodies are super-Earths by investigating the likelihood that a super-Earth projectile may lose its mantle in a hit-and-run collision onto a sub-Neptune target. For different collision velocities, we compare the outcome of impacts of a super-Earth projectile onto a sub-Neptune with a 1.67 $M_\oplus$ iron core, 3.33 $M_\oplus$ rocky mantle and 1.25 $M_\oplus$ hydrogen atmosphere to the predictions for the same collisions assuming that the target is a larger super-Earth \citep[the predictions are based on the SPH simulations from][and our machine-learning model of giant impacts, respectively]{2022MNRASDenman}. The results are in Figure \ref{fig:volatile-rich}a, which plots the density of the surviving hit-and-run projectile ($\rho_S$, in units of $\bar{\rho}$) as a function of $v_{coll}$ for the two different scenarios with projectile mass $M_P$ = 1 and 3 $M_\oplus$. Over the same range of $v_{coll}$, we find that $\rho_S/\bar{\rho}$ is above the high-density threshold (dashed red line, Eq. \ref{eq:high-density_treshold}) if the target is a super-Earth, while it is below the threshold if the target is a sub-Neptune. This suggests that erosion of mantle materials in impacts onto atmosphere-rich sub-Neptune-sized planets is less efficient than direct impact onto solid super-Earth-sized targets.

Higher impact energies than those achieved in collisions between a super-Earth and a sub-Neptune could be achieved if a super-Earth collides onto a Jupiter-sized planet. However, formation of a metal-rich world in this scenario is unlikely, for two reasons. First, no observed high-density exoplanet has a detected Jupiter-sized companion (Figure \ref{fig:HDsystems}). Second, as the projectile-to-target radius ratio decreases, the super-Earth projectile would need to graze the target at $\theta>$ 60$^\circ$ to be stripped of its mantle and avoid being accreted by the Jupiter-sized planet (Figure \ref{fig:volatile-rich}b). However, the most likely impact angle is $\theta$ = 45$^\circ$ and there is only $\sim$ 25\% probability that $\theta>$ 60$^\circ$ (Section \ref{sec:initial_conds} and Figure \ref{fig:input_conditions}d).

Overall, we conclude that a super-Earth is more likely to evolve into, and survive as, a metal-rich world in a hit-and-run collision if the target is a similar-sized super-Earth.

\subsection{Effect of parent-body size on mantle stripping}
\label{sec:size_barrier}

\begin{figure*}[p]
	\centering
 \includegraphics[width=0.95\textwidth]{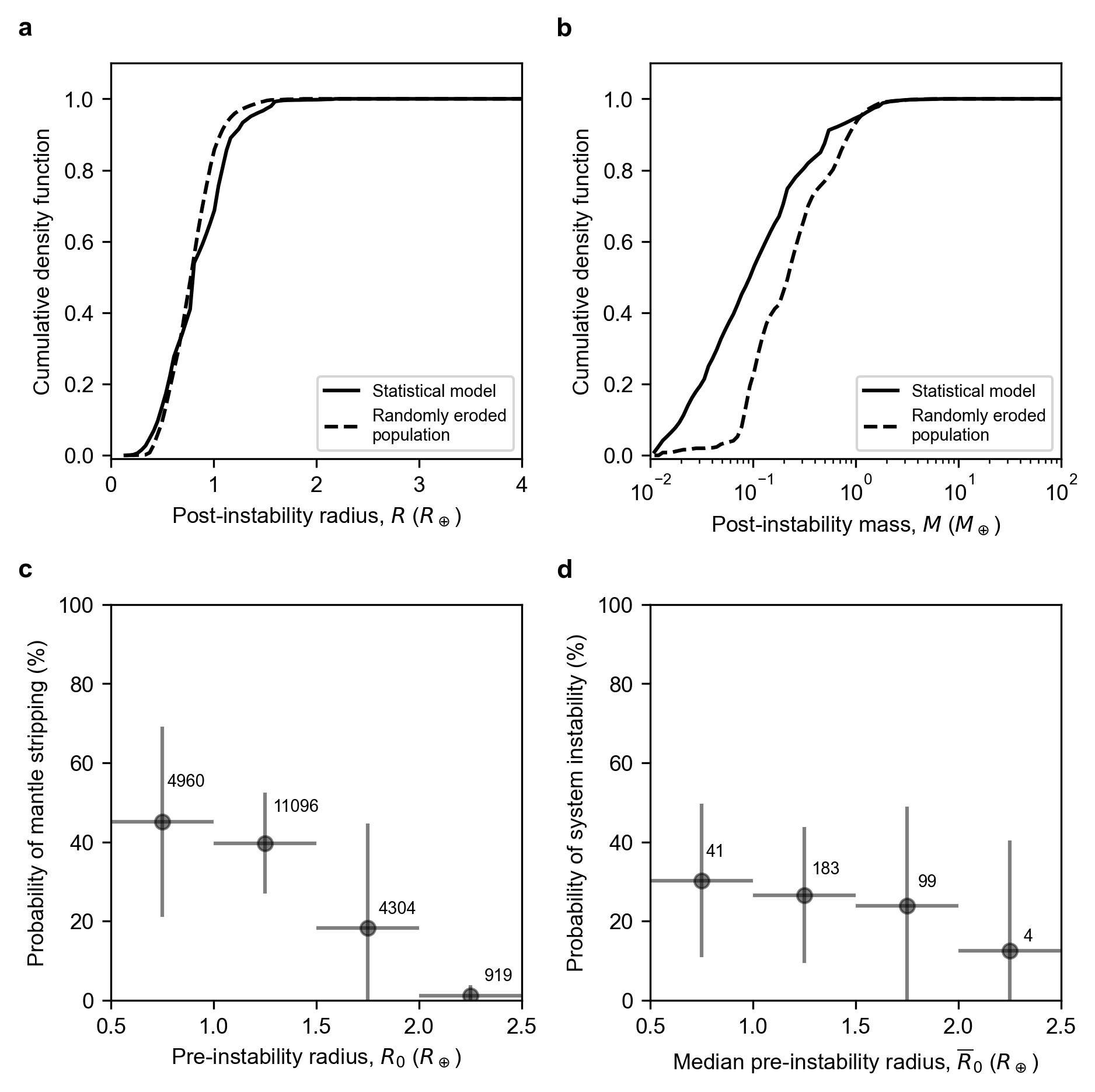}
	\caption{\textbf{The likelihood that mantle-stripping giant impacts occur decreases with increasing size of the colliding bodies}. \textbf{a}, Cumulative distribution of the radii, $R$, of the simulated metal-rich planets compared to that of the corresponding randomly eroded population generated as described in Section \ref{sec:size_barrier}. \textbf{b}, Same as panel a, but for planetary masses, $M$. \textbf{c}, Fraction of initial super-Earths in unstable compact systems that evolve into metal-rich planets (i.e., probability of mantle stripping) as a function of their initial size, $R_0$. \textbf{d}, Fraction of compact systems that experienced instability (i.e., probability of system instability) as a function of the median initial planetary size, $\overline{R}_0$. The value of $\overline{R}_0$ is representative of the typical size of planets in a compact system within 0.14 $dex$ \citep[1 $dex$ = 1 order of magnitude,][]{2022PPVIIWeiss}. In panels c and d, vertical bars indicate 1 standard deviation uncertainties computed across all simulation scenarios described in Table \ref{tab:results_compact} and the annotated numbers indicate the total number of planets and systems, respectively, in each bin.  The widths of the bins are indicated by the horizontal errorbars.}
	\label{fig:trend}
\end{figure*}

The radius of the simulated metal-rich planets, $R$, ranges from $R \approx 0.2 R_\oplus$, smaller than planet Mercury and GJ 367 b, to $R \approx 2.2 R_\oplus$, larger than all observed high-density exoplanets listed in Table \ref{tab:metal_rich_Pgi} (Figures \ref{fig:all_MR_KDE} and \ref{fig:all_MR}). This indicates that mantle-stripping giant impacts between super-Earths can form metal-rich worlds as large as the observed high-density exoplanets.

Nevertheless, we also find that most simulated metal-rich planets are Earth-sized or smaller, with a 95th percentile of the radius distribution, $R_{95}$, of 1.2 $R_\oplus$ for the nominal scenario. Notably, the value of $R_{95}$ varies from 1.0 to 1.6 $R_\oplus$ across the different scenarios considered (a summary of the results is in Table \ref{tab:results_compact}). The minimum $R_{95}$ is recorded for the scenarios in which super-Earths have a lower metal content than the nominal scenario (T2). The maximum is recorded for the scenarios in which super-Earths have a higher metal content than the nominal scenario (T3) or if the precursor systems have super-Earths with $R_0 = 2\text{--}3 R_\oplus$ (T6). Assuming that the nominal scenario and its variations are all equally probable, we average the 109,000 simulation results and obtain $R_{95} = 1.3 \pm 0.2 R_\oplus$, where the uncertainty corresponds to 1 standard deviation.

We investigate whether the preferentially sub-Earth size of the simulated metal-rich worlds is due to their parent bodies being super-Earths with $R_0< 2\text{--}3 R_\oplus$ (Figure \ref{fig:input_conditions}a). To test this, for a given super-Earth of size $R_0$, mass $M_0$, and core-mass fraction $Z_0$ that becomes metal-rich in at least one simulation from our statistical model, we randomly sampled $Z$ between $Z_0$ and 1 and estimated the corresponding mass as $M\sim Z_0/Z \times M_0$. We find that the simulated metal-rich planets have indeed a radius distribution that is consistent with that of this randomly eroded population (Figure \ref{fig:trend}a). This indicates that the size of the simulated metal-rich planets is indeed limited, at least in part, by that of their parent bodies.

However, we also find that the simulated metal-rich planets from the model of orbital instabilities tend to have a lower mass, $M$ (or equivalently, a lower density for a given $R$), than the metal-rich planets of the randomly eroded population (Figure \ref{fig:trend}b). This suggests that the size distribution of the simulated metal-rich worlds is also affected by the binding energy of their parent super-Earths, which opposes their impact disruption \citep[e.g.,][]{2016IcarusMovshovitz}. Because binding energy increases with increasing planetary size \citep[e.g.,][]{2023AREPSGabrielCambioni}, the percentage of mantle erosion is expected to decrease with increasing planetary size for a given range of $v_\infty$. Consistent with this, we find that the probability that a super-Earth suffers mantle stripping decreases with $R_0$ (Figure \ref{fig:trend}c), dropping from ${\sim}50\%$ to ${\lesssim} 5\%$ for $R_0 < 1 R_\oplus$ and $R_0> 2 R_\oplus$, respectively.

In addition, astronomical observations suggest that the spacing between adjacent planets in compact systems increases with planetary size \citep{2021AAMishra,2022PPVIIWeiss} and that their mutual eccentricities are generally smaller than 0.1 \citep{2017AJHadden}. Because the likelihood of instability tends to decrease with increasing inter-planetary distances \citep{2015ApJPuWu}, compact systems with larger planets may be inherently less likely to experience instability than systems with smaller planets. Consistent with this, in our results, we find that the fraction of precursor systems that experience instability appears to decrease with the median value of the initial size of their planets, $\overline{R}_0$ (Figure \ref{fig:trend}d), although less markedly than how the probability of mantle stripping decreases as a function of $R_0$ (Figure \ref{fig:trend}c). As such, we consider the decrease in the likelihood of compact-system instability as a function of increasing average size of the planets in the compact systems to have a secondary effect in shaping the size of the metal-rich giant-impact remnants.

\subsection{Implications for the origin of high-density exoplanets} \label{sec:datavsmodel}

\begin{figure}[p]
	\centering
	\includegraphics[width=0.9\linewidth]{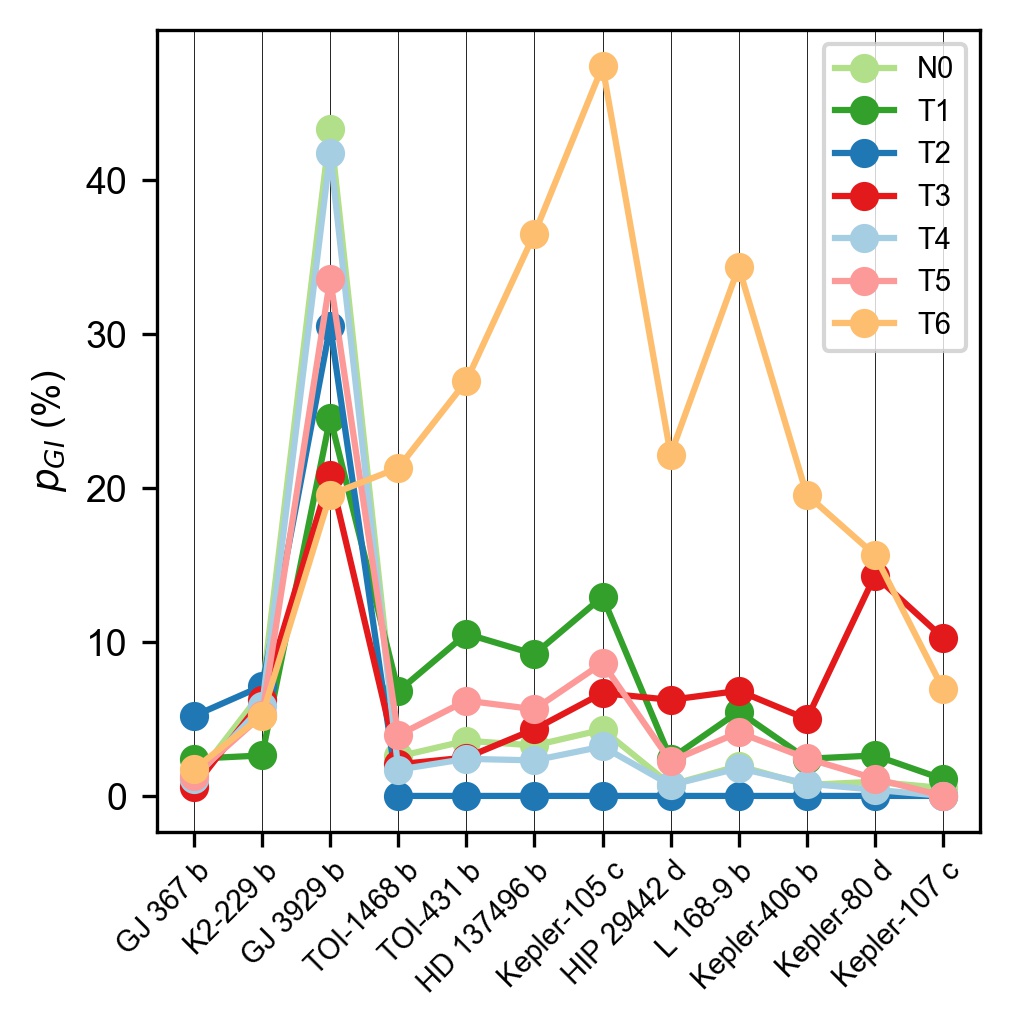}
	\caption{\textbf{Probability $p_{GI}$ that the orbital instability forms metal-rich worlds as large and massive as the observed high-density exoplanets}. The probability is computed as described in Section \ref{sec:exosurvey} and the planets are listed along the x-axis in order of increasing size.}
	\label{fig:Pgi}
\end{figure}

Our results demonstrate that super-Earths may become metal-rich worlds by losing their mantles in giant impacts. Nevertheless, we also find that the formation of metal-rich giant-impact remnants is a rare occurrence, with a predicted observed occurrence rate among the terrestrial exoplanets in the post-instability population of $\tilde{\eta}_{MR} = 0.5 \pm 0.5$, where the uncertainty is 1 standard deviation (Eq. \ref{eq:predicted_MR} in Section \ref{sec:eta_MR}; Table \ref{tab:results_compact} and Figure \ref{fig:all_MR}). This is much lower than the occurrence rate of the observed high-density exoplanets ($\eta_{HD} = 9\%$, Section \ref{sec:observed_occurrence_rate}). When scaled to the number of observed terrestrial exoplanets, $\tilde{\eta}_{MR}$ translates in at most 1 observed high-density exoplanets being a metal-rich giant-impact remnant, against 12 observed high-density planets. Importantly, our estimate of $\tilde{\eta}_{MR}$ is most likely an upper limit, for two reasons. First, we are overestimating the number of metal-rich worlds by assuming that all unstable compact systems will undergo a phase of giant impacts. Second, we are underestimating the total number of terrestrial exoplanets by considering only exoplanets in the post-instability populations. In reality, not all unstable systems would undergo giant impacts to achieve stability and not all systems with terrestrial exoplanets would experience an instability.

Following the procedure outlined in Section \ref{sec:KS_test}, we find that the mass-radius distribution of the simulated metal-rich worlds corrected for observational biases has KS probability smaller than 5$\times$10$^{-15}$ to be drawn from the same underlying distribution as that of the observed high-density exoplanets. This mismatch can be explained by most of the observed high-density exoplanets being larger and more massive than the simulated metal-rich worlds (Section \ref{sec:size_barrier}). 
Consistently, we find that all scenarios (except T6, discussed below) form metal-rich worlds as large and massive as the observed high-density exoplanets only in $p_{GI}<$ 15\% of the cases (Figure \ref{fig:Pgi}). 
The exception is the observed high-density exoplanet GJ 3929 b, for which $p_{GI} = 20\text{--}40\%$, depending on the scenario. This may be explained by GJ 3929 b having the lowest probability of being high-density among the observed high-density exoplanets (Table \ref{tab:metal_rich_Pgi}), making it easier to form via mantle stripping starting from a $Z_0$ between 0.2 and 0.4.

The exception to the trend in $p_{GI}$ described above is scenario T6, that is, the scenario in which we explore the orbital instability of compact systems that have planets with $R_0 = 2\text{--}3 R_\oplus$, thus larger than the typical maximum size of $2 R_\oplus$  adopted in the nominal scenario. This scenario yields $p_{GI} > 20\%$ for the high-density exoplanets of sizes between GJ 3929 b and Kepler-80 d (that is, from ${\sim}1 R_\oplus$ to $1.5 R_\oplus$), with a maximum $p_{GI}\sim  47\%$ for Kepler-105 c. We explain the presence of a peak in $p_{GI}$ (Figure \ref{fig:Pgi}) as being due to ${\sim}90\%$ of the initial super-Earths in the scenario T6 being larger than $1~R_\oplus$ (Figure \ref{fig:input_conditions}b). The lack of sub-Earth-sized planets in the precursor systems results in reduced collisional grinding at small scales. At the same time, the largest super-Earths in this scenario are too large for giant impacts to efficiently erode their mantles (Figure \ref{fig:trend}c). The competing factors of lack of bodies to grind down at small scales and reduced collisional grinding at large scales create a peak in the abundance of simulated metal-rich worlds at an intermediate size. Intriguingly, the 1-D probability density functions of masses and radii of the simulated metal-rich planets from scenario T6 visually resemble those of the observed high-density exoplanets (Figure \ref{fig:all_MR_KDE}). This makes scenario T6 the best-fit scenario among those tested to form metal-rich worlds akin to the observed high-density exoplanets.

Despite this, the formation of metal-rich worlds in scenario T6 remains a rare event, yielding a predicted observed occurrence rate $\tilde{\eta}_{MR} < 1.4\%$ among terrestrial planets (Table \ref{tab:results_compact} and Figure \ref{fig:all_MR}), thus lower than the occurrence rate of observed high-density exoplanets. Moreover, scenario T6 does not pass the KS test, neither in the 2-dimensional (mass-radius) space nor if the 1-D probability mass and radius density functions of the simulated versus modelled planets are separately compared. Finally, the compact systems of scenario T6 are relatively rare, accounting for ${\sim}14\%$ of all compact systems in the catalog (Appendix \ref{appx:T6_occurrence}), although it is possible that more of these systems could have existed in the past than are observed today. As such, we conclude that it is unlikely that scenario T6 can explain all the observed high-density exoplanets, although we cannot rule out that it may explain some of the observed high-density exoplanets of size $1\text{--}1.5 R_\oplus$.

\section{Conclusion}
\label{sec:conclusion}
High-density worlds are a recently-recognized compositional class of planetary objects whose density is far higher than what expected from cosmochemistry and whose sizes range from metal-rich asteroids to super-Earth-sized planets. In this work, we tested the hypothesis that the largest members of this population (the high-density exoplanets) are the metallic cores of super-Earths that lost their rocky mantles by giant impacts.

To test this hypothesis, we combined exoplanetary observations, orbital dynamics, collisional physics, and machine learning to develop and run a statistical model of orbital instability and collisional evolution of super-Earth systems. In our simulations, we maximized the impact energies (and thus the likelihood of mantle stripping) by studying the case in which the orbital instabilities involve close-in super-Earths that are tightly packed in compact systems around a host star. We explored a comprehensive set of scenarios (109,000 different coupled orbital-collisional simulations) by varying the initial composition, mass, and the degree of orbital excitation of the super-Earths in about 100 observed compact systems. Next, we compared the simulation results in terms of mass, radii, and occurrence rate of simulated metal-rich worlds to those of the observed high-density exoplanets.

We found that metal-rich worlds as massive and large as the observed high-density exoplanets can form as a result of giant impacts between super-Earths during a late orbital instability. The formation mechanism is hit-and-run collisions, in which the metallic core of a projectile super-Earth manages to escape accretion onto a larger super-Earth target. In this process, the mantle of the projectile is blasted off and is sequestered by the larger super-Earth target. This scenario is similar to what is proposed for planet Mercury \citep[e.g.,][]{2014NatGeoAsphaug}.

Nevertheless, we also find that most giant impacts occurring during a late instability result in pairwise accretion of the colliding super-Earths, yielding a predicted observed occurrence rate for the simulated metal-rich worlds among terrestrial planets smaller than ${\sim}1\%$. By contrast, we estimate that the observed occurrence rate of high-density exoplanets is ${\sim}9\%$. In addition, the mass-radius distribution of the metal-rich worlds is statistically distinct from that of the observed high-density exoplanets, with Kolmogorov-Smirnoff probability smaller than $5 \times 10^{-15}$. Consistently, at most 15\% of the metal-rich worlds formed during instabilities are as large and massive as the observed high-density exoplanets. This conclusion is largely insensitive to our model assumptions (Section \ref{subsec:model_variations}). The only exception is the scenario in which the parent bodies of the metal-rich worlds are rare super-Earths of terrestrial composition with radius of $2\text{--}3 R_\oplus$, intermediate between the radius typical of super-Earths and sub-Neptunes. The instability of these rare systems may produce larger metal-rich planets than in other scenarios, albeit still with a predicted observed occurrence rate in the order of 1\%. %

We conclude that mantle-stripping giant impacts between super-Earths can form metal-rich worlds as large as the observed high-density exoplanets,  but these events occur at rates not sufficient to explain the typical size and abundance of the observed high-density exoplanets. Based on the current observed rate, this translates into at most one of the observed high-density exoplanets likely to be a metal-rich giant-impact remnant. This planet may be GJ 3929b, whose probability to be a metal-rich giant impact remnant is the highest, 20-40\% depending on the scenario. If future exoplanetary surveys were to confirm the current high-density exoplanet occurrence rate, then there would be only two remaining hypotheses likely to explain the detection of most high-density exoplanets: these planets are either primordial metal-rich worlds, the naked compressed cores of gas giants that lost their volatile envelopes, or a mixture of these two populations \citep{2022AAJohansen,2014RSPTAMocquet}. If future surveys would instead lower the occurrence rate down to less than 1--2\%, our results suggest that mantle-stripping giant impacts may still be a viable formation mechanism.

Finally, our simulations yield an occurrence rate of metal-rich giant-impact remnants smaller than Earth of 21 $\pm$ 12\% among terrestrial exoplanets ($1\sigma$ uncertainty). As such, we predict that sub-Earth-sized metal-rich worlds should be relatively common in the galaxy but are likely too small to be detected by current exoplanetary surveys. The formation rate of these planets ranges between $\sim$9 and $\sim$43\% across all our model realizations that assume an abundance of iron and rock-forming elements of the protoplanetary disk as that of CI carbonaceous chondrites. This finding supports the proposal that the planet Mercury and some metal-rich asteroids are metal-rich worlds that formed by giant impacts \citep{2014NatGeoAsphaug, 2015ApJVolkGladman, 2022SSRvElkins}, without the need for a metal enrichment of the local protoplanetary disk.\\

\textbf{Acknowledgements}: S. C. acknowledges the Crosby Distinguished Postdoctoral Fellowship of Department of Earth, Atmospheric and Planetary Sciences, Massachusetts Institute of Technology (MIT). B. P. W. acknowledges funding from NASA contract NNM16AA09, ``Psyche: Journey to a Metal World.'' K. V. acknowledges support from NASA (Exoplanets Research Program grant 80NSSC18K0397 and ``Alien Earths" grant 80NSSC21K0593); the results reported herein benefited from collaborations and/or information exchange within NASA’s Nexus for Exoplanet System 
Science (NExSS) research coordination network sponsored by NASA’s Science Mission Directorate. Z. L. acknowledges funding from the Center for Matter at Atomic Pressures (CMAP), a National Science Foundation (NSF) Physics Frontiers Center, under Award PHY-2020249.
The authors acknowledge the MIT \href{https://supercloud.mit.edu/}{SuperCloud} and Lincoln Laboratory Supercomputing Center for providing high-performance computing, database, and consultation resources. 

\clearpage
\begin{appendix}
\counterwithin{figure}{section}
\counterwithin{table}{section}
\section{Appendix}

\subsection{Estimation of the effect of the RV bias on $\eta_{HD}$}
\label{appx:RV_bias}

Most of the terrestrial exoplanets in our catalog ($\sim$90\%) were discovered with the transit method. In most cases, their discovery was later followed by mass measurements using the radial velocity (RV) method. The signal semi-amplitude of RV measurements, $K$, increases with $M$ as
\begin{equation}
\label{eq:RV_amp}
    K = \bigg(\frac{2\pi G}{P}\bigg)^{\frac{1}{3}} \frac{M\sin I}{(M+M_*)^{\frac{2}{3}}}\frac{1}{\sqrt{1-e^2}}
\end{equation}
\noindent
where $M_*$ is the host star mass and $I$ is the angle between the plane in which the planet orbits and the plane perpendicular to the line of sight. Because the mass of a high-density exoplanet is larger than that of a similarly-sized planet with an Earth-like composition, the former has higher $K$ than the latter assuming a similar orbital period. This effect may make it easier to detect or obtain better mass constraints on the high-density exoplanets than of terrestrial planets, potentially leading to an overestimation of $\eta_{HD}$ as computed in Section \ref{sec:observed_occurrence_rate}.

Quantifying the correction factor for $\eta_{HD}$ to account for this bias is not straightforward because $P$, $M$, $M_*$ (and to a lesser extent, $I$ and $e$) are highly variable across exoplanets. Furthermore, these values are a priori unknown for planets whose RV signal is below current detection limits. Therefore, here we limit ourselves to bound the effect of the RV bias on $\eta_{HD}$ through two worst case scenarios.

First, we assume that the RV bias always leads to larger uncertainties in the RV masses for less massive planets. To explore this, we redo our search for terrestrial exoplanets as described in Section \ref{sec:observed_occurrence_rate}, but assuming a more permissive $\sigma_M/M< 0.5$ instead of $\sigma_M/M < 0.3$. We still use the criteria $\sigma_M/M < 0.3$ for the high-density exoplanets. Second, we assume that the RV bias always leads to missed detections of terrestrial planets with an Earth-like composition, but never of similarly-sized high-density exoplanets. To explore this, we build a synthetic population of 100 terrestrial exoplanets with $Z = 0.3$ and radii randomly sampled from the radius distribution of the high-density exoplanets in Table \ref{tab:metal_rich_Pgi}. We assume that the masses of these terrestrial planets can be constrained by the RV method if their $K$ as defined by Eq. \ref{eq:RV_amp} is greater than an assumed RV detection limit $K_{th}$. We explore how the number of detected terrestrial planets, $N_R$, varies as a function of $K_{th}$ between $\sim$0.3 cm/s and 10 m/s. The lowest value $K_{th}$ is the best precision level achieved by the latest generation of spectrographs, while the highest value of $K_{th}$ is $\sim$100 times the velocity variation that the Earth produces on the Sun \citep[][for a review]{2023AnRSAHara}. We assume $I=$ 90$^\circ$, $e$ = 0, $M_*$ equal to one solar mass for all planets, and explore $P$ = 1 day, 10 days, or 100 days.

\begin{figure}
	\centering
	\includegraphics[width=\linewidth]{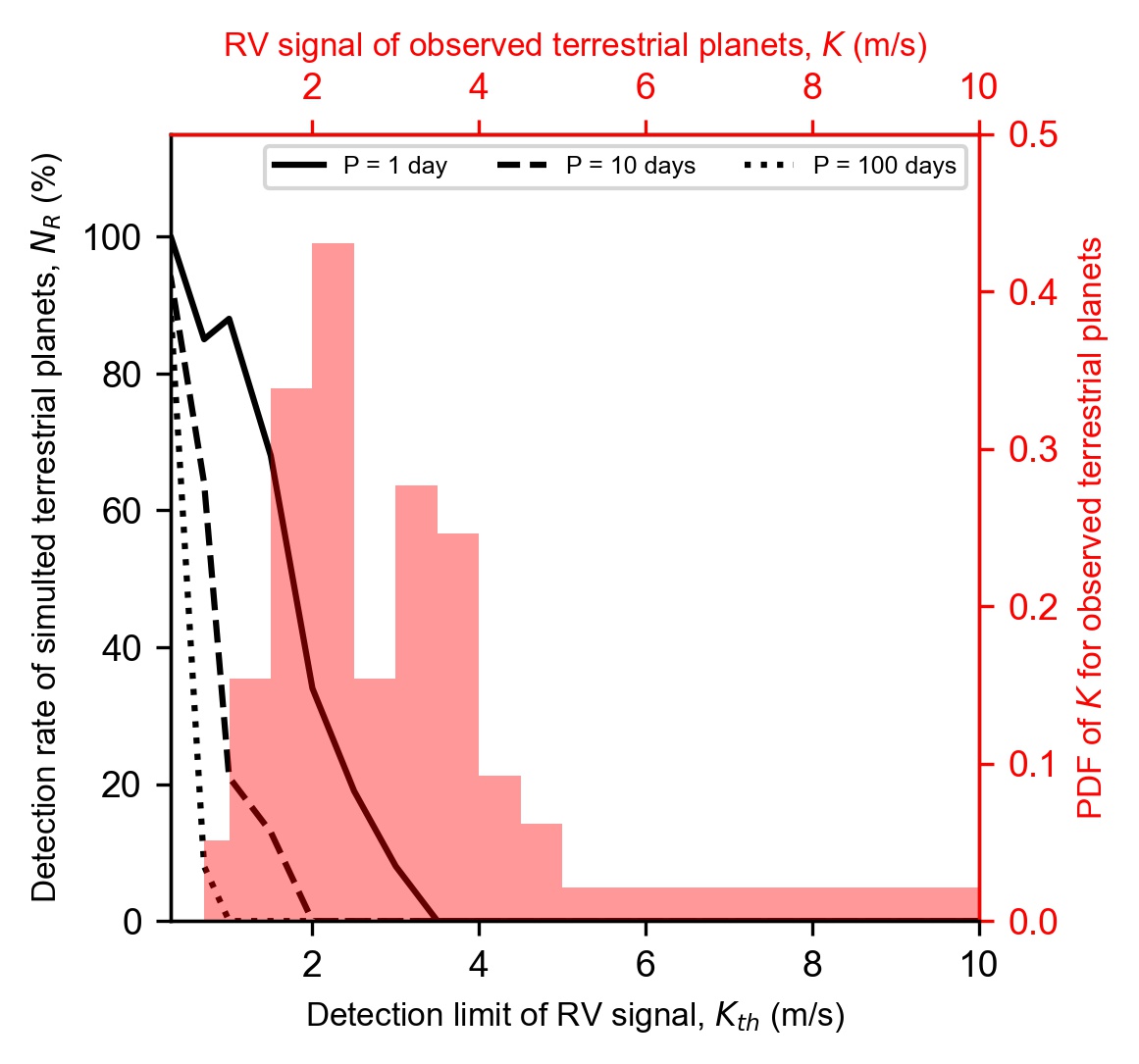}
	\caption{\textbf{RV bias effect on the detection of terrestrial planets.} Left-bottom axes: Percent of terrestrial exoplanets detected, $N_R$, as a function of the detection limit of RV signals, $K_{th}$, for different values of the orbital period $P$ of the 100 terrestrial exoplanets in the model catalog (see Appendix \ref{appx:RV_bias}). Right-top axes: Probability Density Function (PDF) of the RV signals of the observed terrestrial exoplanets in our dataset.}
	\label{fig:RV_bias}
\end{figure}

The first test yields $\eta_{HD} = 12/103 = 12\%$ assuming that all terrestrial planets have $R < 2 R_\oplus$. 
In the second test, we find (Figure \ref{fig:RV_bias}) that $N_R$ drops from 100\% to 0\% for $K_{th}>$ 3.5, 2.0 and 1 m/s if $P$ = 1, 10, and 100 days respectively. We note that 54\%, 21\% and 2\% of the observed terrestrial exoplanets in our dataset have  $K>K_{th}$ = 3.5, 2.0 and 1 m/s, respectively (histogram in Figure \ref{fig:RV_bias}). We therefore infer that at most 54\% of the terrestrial planets may not have been detected due to the RV bias. This translates in a minimum $\eta_{HD}$ of 9\% assuming that the 12 high-density exoplanets are so massive to be detectable regardless of the value of $K_{th}$. The unbiased value of $\eta_{HD}$ is still at least 9 times larger that the predicted observed occurrence rate of the simulated metal-rich worlds $\eta_{MR}<$ 1\% (Section \ref{sec:datavsmodel}). We therefore conclude that our results presented in the main text are robust to the RV bias as quantified here.

\subsection{Occurrence rate of compact systems for scenario T6}
\label{appx:T6_occurrence}
We estimate the occurrence rate of the compact systems in scenario T6 over the entire catalog of compact systems (thus, including both the compact systems from scenarios T6 and N0) as:
\begin{equation}
    \eta_{sys} = \frac{N_{T6}~(1+f_{T6})}{N_{T6}~(1+f_{T6})+{N_{N0}~(1+f_{N0})}}
\end{equation}
where $N$ is the number of compact systems and $f$ indicates the instability rate that we estimated for the compact systems. We use the latter to correct for those systems that are no longer compact because they went unstable in the past. For the scenario T6, Table \ref{tab:results_compact} yields $N_{T6}$ = 12, $N_{N0}$ = 90, $f_{T6}$ = 0.5, $f_{N0}$ = 0.23. This translates to $\eta_{sys}$ = 14\%, revealing that the compact systems in scenario T6 are relatively rare among the whole observed set of compact systems.

\begin{figure*}[hbt!]
	\centering
	\includegraphics[width=\textwidth]{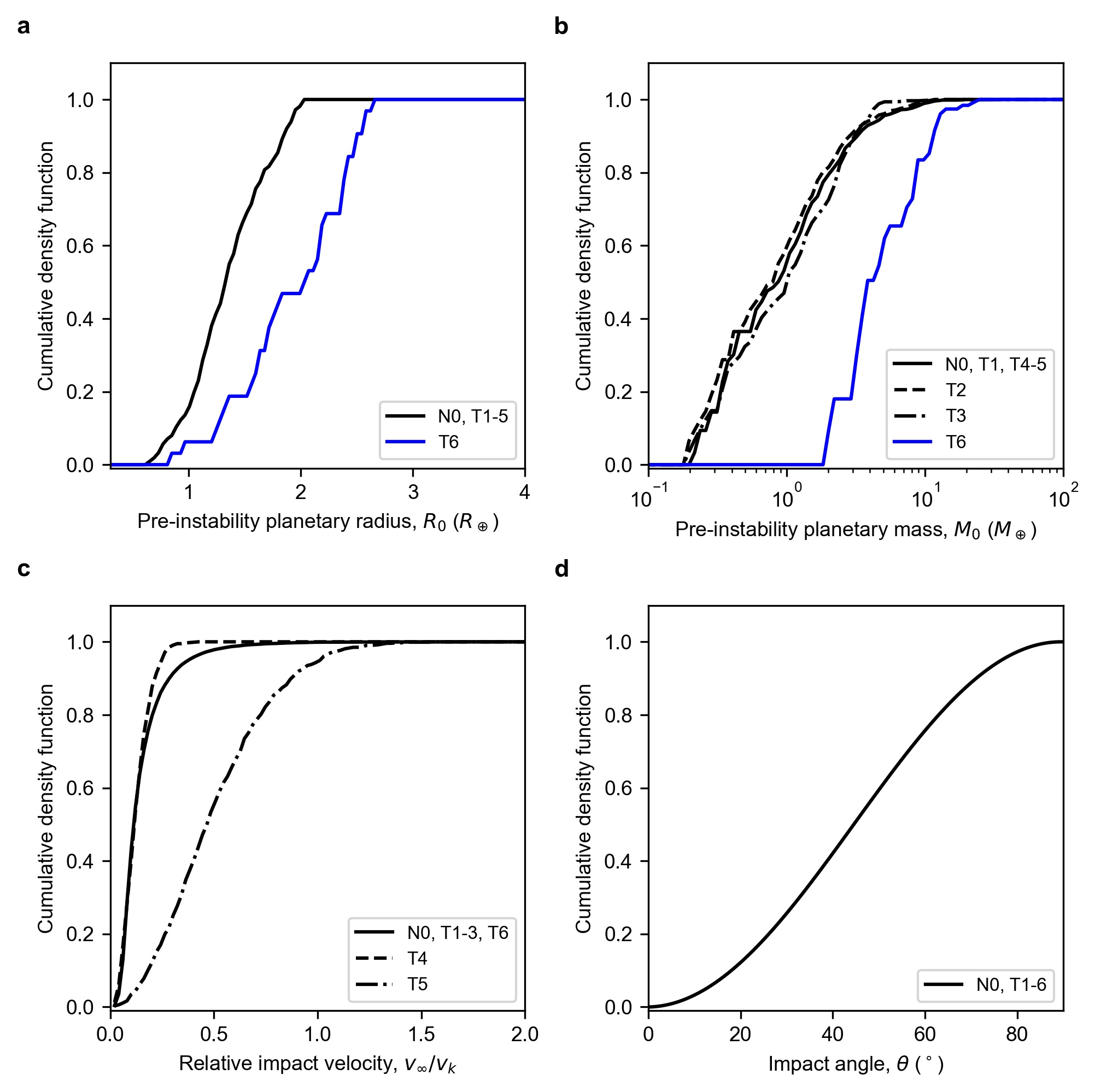}
 \caption{\textbf{Input probability distributions for the orbital statistical model}. \textbf{a}, Distribution of initial planets' radius, $R_0$, in the precursor compact systems. \textbf{b}, Distribution of initial planets' mass, $M_0$, in the compact systems.  
 \textbf{c}, Distribution of the relative impact velocity in units of Keplerian velocity, $v_{\infty}/v_{k}$ assumed in the statistical model of Section \ref{sec:statistical_model}. \textbf{d}, Distribution of the impact angle, $\theta$ assumed for the collision model in Section \ref{sec:statistical_model}. The different curves are for the nominal scenario (N0) and its variations (T1--T6).}
	\label{fig:input_conditions}
\end{figure*}

\begin{figure*}[]
	\centering
	\includegraphics[width=0.80\textwidth]{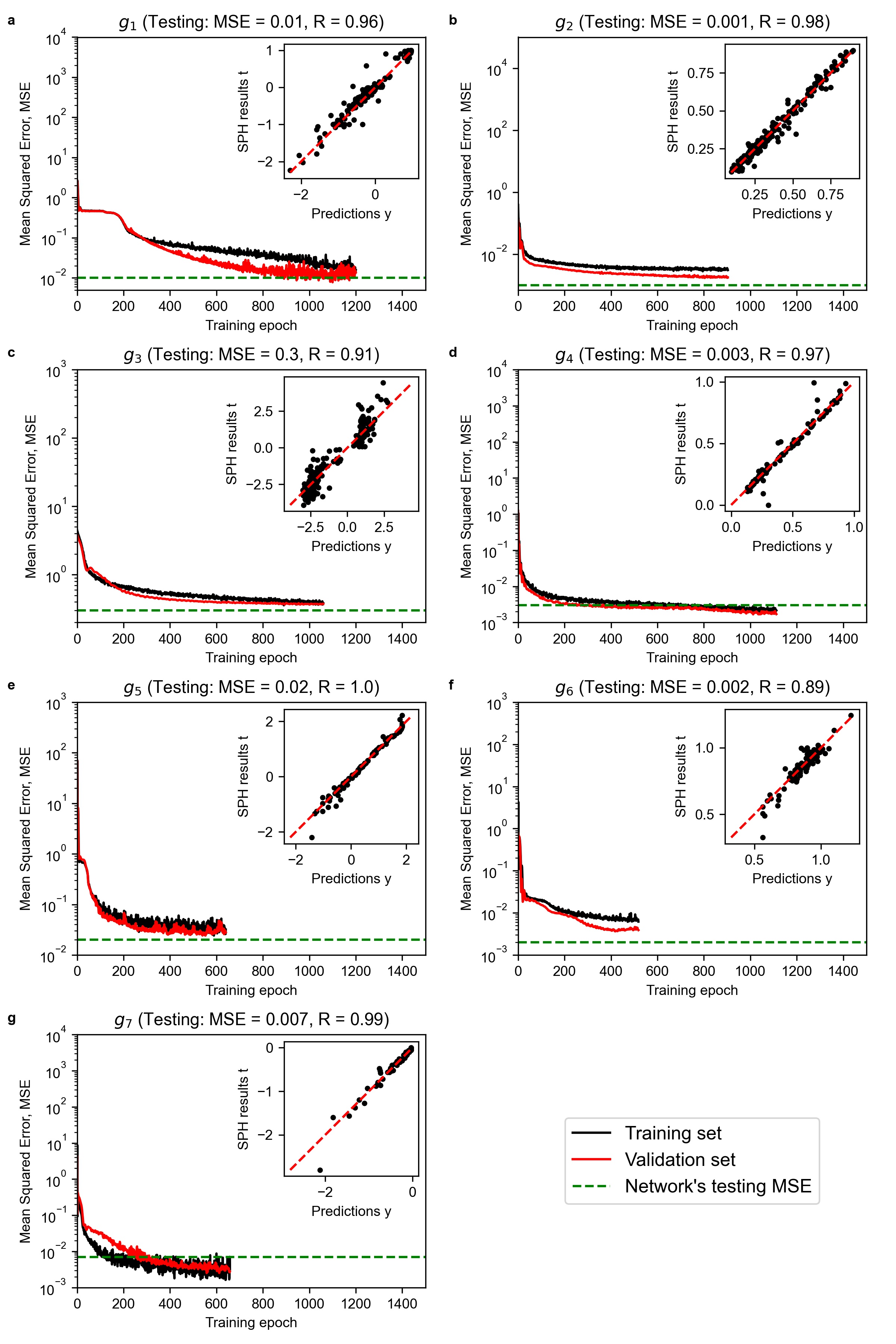}
	\caption{\textbf{Training, validation and testing of the optimal neural networks composing the giant-impact model}. \textbf{a--g}, mean squared error (MSE) between $y(x)$ predicted by the neural networks $g_i$ ($i$ = 1,...,7) and the corresponding SPH outcome, $t(x)$, as a function of the training epochs. The inset plots show the correlation with index R between $y(x)$ and $t(x)$ for the testing set.}
	\label{fig:NNPerformance}
\end{figure*}

\begin{figure}
    \centering
\includegraphics[width=\linewidth]{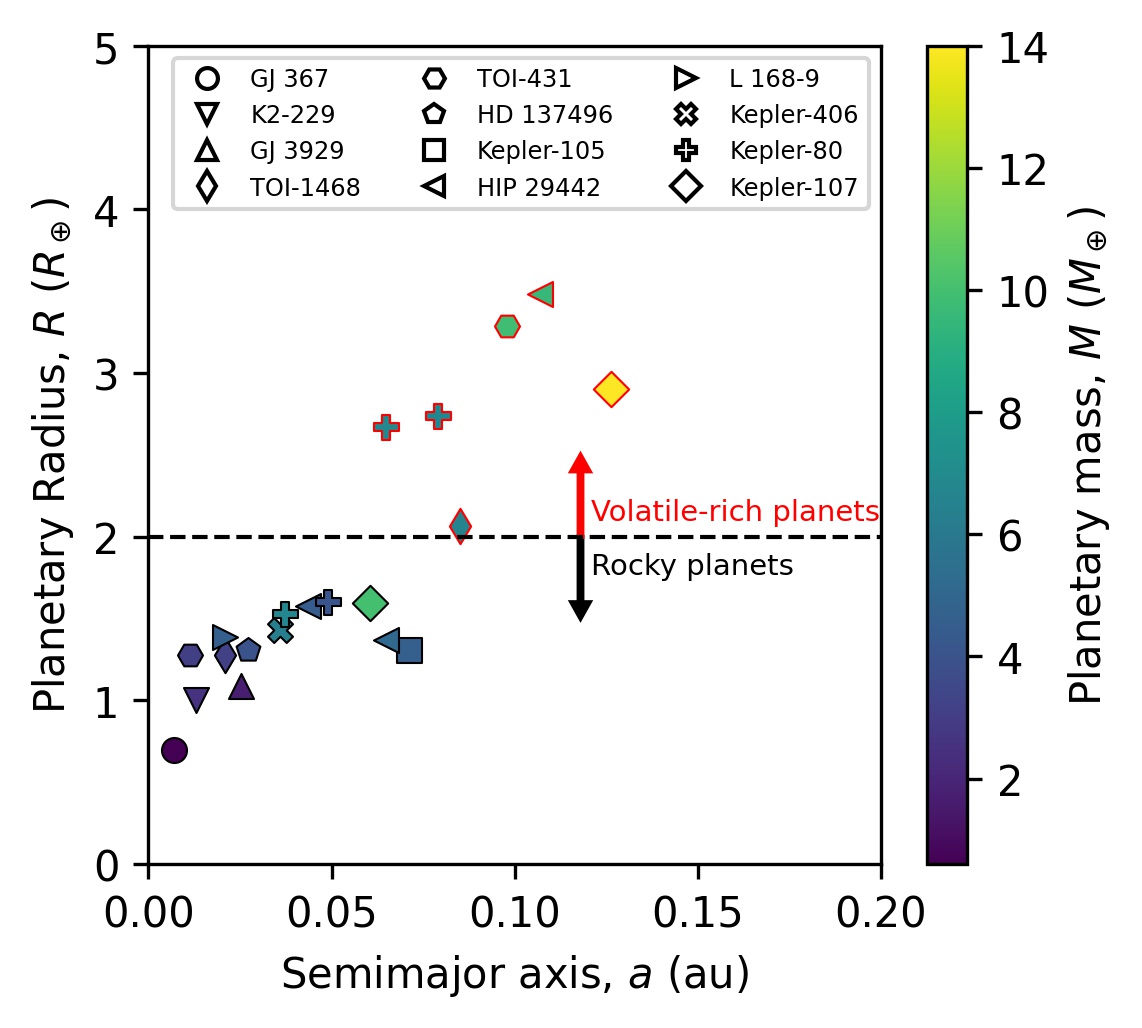}
    \caption{\textbf{Most of the known planets in the systems with high-density exoplanets are made of rock and iron, with no large volatile envelopes}. Radius, $R$, as a function of semimajor axis, $a$, of planets in the planetary systems where high-density exoplanets have been observed, color coded by their masses, $M$ (in units of $M_\oplus$). We plot only planets for which the radius and masses are known with sufficient precision to enable compositional estimates, i.e., $\sigma_R/R<$0.3 and $\sigma_M/M<$0.3. We distinguish between terrestrial planets (symbols with black edges) and volatile-rich planets (symbols with red edges) based on whether their $R$ is greater or smaller than 2 $R_\oplus$, respectively (see Section \ref{sec:observed_occurrence_rate}).}
	\label{fig:HDsystems}
\end{figure}

\begin{figure}
    \centering
\includegraphics[width=\linewidth]{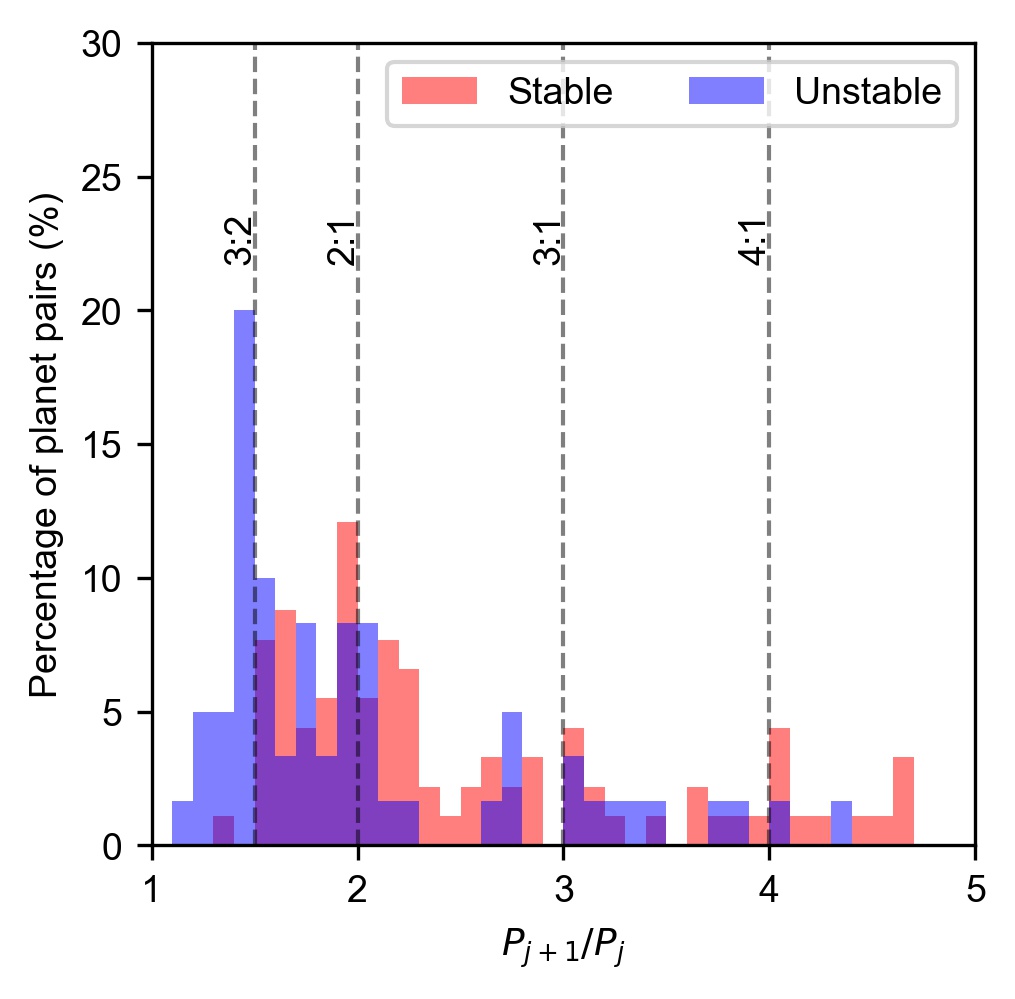}
    \caption{\textbf{Period ratios of planets in the precursor compact systems.} The red and blue histograms concern planets in compact systems that we find to be stable and likely to become unstable within the next 10-100 Myr, respectively, following the procedure outlined in Section \ref{sec:statistical_model}. In the stable systems, we find that 30\% of the planetary pairs are within $\pm$0.1 $P_{j+1}/P_{j}$ of mean motion resonances (7\%, 15\%, 4\% and 5\% have period ratios 3:2, 2:1, 3:1, 4:1, respectively). In the systems that we predict to be unstable, we find that 50\% of the planetary pairs are within $\pm$0.1 $P_{j+1}/P_{j}$ of mean motion resonances (28\%, 16\%, 3\% and 2\% have period ratios 3:2, 2:1, 3:1, 4:1, respectively). Overall, this translates in 37\% of the planetary pairs in the compact systems being within $\pm$0.1 $P_{j+1}/P_{j}$ of the 3:2, 2:1, 3:1 and 4:1 mean motion resonances. System-wise, 20\% of the stable systems have planets in near--mean-motion resonances, while 70\% of the unstable systems have planets in near--mean-motion resonances.}
	\label{fig:MMRs}
\end{figure}

\begin{figure}[t]
	\centering
	\includegraphics[width=\linewidth]{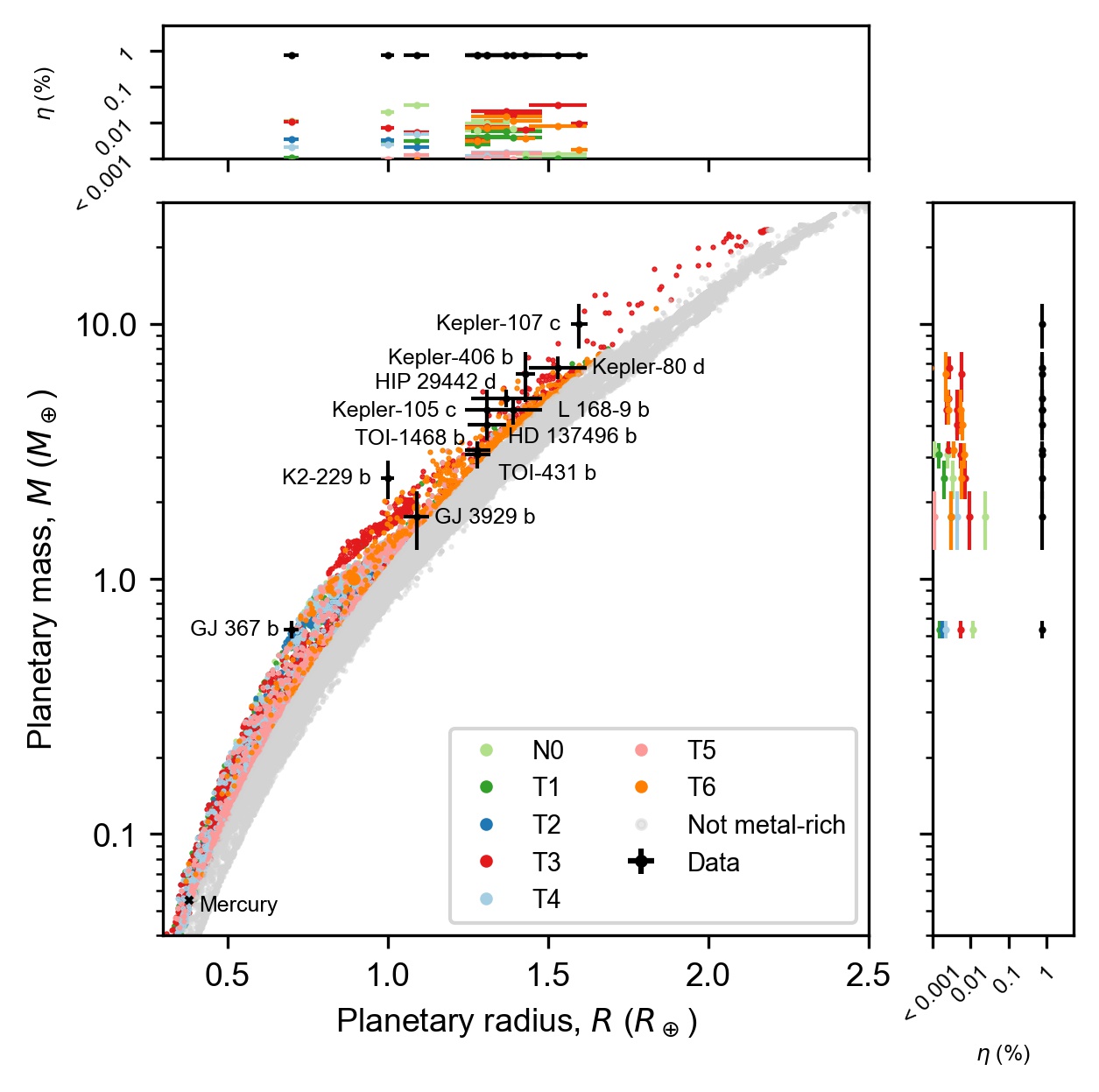}
	\caption{\textbf{Results of the statistical model for different model assumptions}. Mass, $M$, as a function of radius, $R$, of the simulated planets surviving an orbital instability for the different model assumptions described in Section \ref{sec:initial_conds}. The final planets are deemed metal-rich remnants (colored symbols) and non--metal-rich remnants (grey symbols) based on whether their density is higher (lower) than the high-density threshold of Eq. \ref{eq:high-density_treshold}. The black data points correspond to the observed high-density exoplanets listed in Table \ref{tab:metal_rich_Pgi}, with 1 standard deviation uncertainty bars; planet Mercury is plotted for comparison. The side plots show the estimated occurrence rate among terrestrial planets, $\eta$, of the observed high-density planets (black points, which integrates to $\eta_{HD}$, Section \ref{sec:observed_occurrence_rate}) and the simulated metal-rich giant-impact remnants (colored points, which represent the sum of $\tilde{\eta}_{MR}$ within the same 1 $\sigma$ bins of the black data points) as a function of radius (top) and mass (left).}
	\label{fig:all_MR}
\end{figure}

\begin{table*}
\caption{\textbf{Metal-rich planets formed during an orbital instability in the nominal scenario (N0) and its variations (T1--T6)}. The statistics in every column are computed across 109,000 simulations (1000 simulations for each of the 109 unstable systems). The uncertainties in the rightmost column correspond to 1 standard deviation.}
\begin{center}
\begin{tabular}{|lcccccccc|} 
\hline	
\multicolumn{9}{|c|}{\textbf{Statistics of compact systems' instability}} \\
\hline	
\textbf{}	&	\textit{N0}	&	\textit{T1}	&	\textit{T2}	& 	\textit{T3} & \textit{T4} & \textit{T5} & \textit{T6} & \textbf{Average}\\
\hline
Total systems	&	90	&	23	&	90	&	90	&	90	&	90	&	12	&	-			\\
N. unstable	&	21	&	14	&	19	&	17	&	19	&	13	&	6	&	-	\\
N. unstable (\%)	&	23	&	61	&	21	&	19	&	21	&	14	&	50	&	29 $\pm$ 17	\\
	&		&		&		&		&		&		&		&					\\
\hline
\multicolumn{9}{|c|}{\textbf{Pathway of formation of the simulated metal-rich planets}} \\
\hline	
\textbf{}	&	\textit{N0}	&	\textit{T1}	&	\textit{T2}	& 	\textit{T3} & \textit{T4} & \textit{T5} & \textit{T6} & \textbf{Average}\\
\hline
Disruption (\%)	&	0.5	&	0.5	&	3.2	&	1.8	&	0.4	&	0.0	&	0.1	&	1	$\pm$	1	\\
Hit-and-run (\%)	&	99.5	&	99.5	&	96.8	&	98.2	&	99.6	&	100.0	&	99.9	&	99	$\pm$	1	\\
	&		&		&		&		&		&		&		&					\\
\hline
\multicolumn{9}{|c|}{\textbf{Occurrence rate (\%) of the simulated metal-rich planets}} \\
\hline	
\textbf{}	&	\textit{N0}	&	\textit{T1}	&	\textit{T2}	& 	\textit{T3} & \textit{T4} & \textit{T5} & \textit{T7} & \textbf{Average}\\
\hline
${\eta_{MR}}$	&	1.6	&	0.8	&	0.8	&	2.7	&	1.9	&	1.8	&	1.6	&	1.6	$\pm$	0.6	\\
${\eta_{MR}}$ ($R<$1 $R_\oplus$)	&	17.4	&	8.9	&	7.4	&	19.4	&	18.2	&	42.9	&	30.3	&	21	$\pm$	12	\\
${\tilde{\eta}_{MR}}$	&	0.4	&	0.1	&	0.1	&	1.3	&	0.4	&	0.3	&	0.6 &	0.5	$\pm$	0.5	\\
	&		&		&		&		&		&		&		&					\\
\hline
\multicolumn{9}{|c|}{\textbf{Radius $\bm{R~(R_\oplus)}$ of the simulated metal-rich planets}} \\
\hline	
\textbf{}	&	\textit{N0}	&	\textit{T1}	&	\textit{T2}	& 	\textit{T3} & \textit{T4} & \textit{T5} & \textit{T6} & \textbf{Average}\\
\hline
Minimum 	&	0.2	&	0.2	&	0.2	&	0.2	&	0.2	&	0.2	&	0.4	&	0.2	$\pm$	0.1	\\
Mean value	&	0.8	&	0.8	&	0.7	&	0.9	&	0.8	&	0.9	&	1.2	&	0.9	$\pm$	0.2	\\
68th percentile	&	1.0	&	1.0	&	0.9	&	0.9	&	1.0	&	1.0	&	1.3	&	1.0	$\pm$	0.1	\\
95th percentile	&	1.2	&	1.3	&	1.0	&	1.6	&	1.2	&	1.3	&	1.6	&	1.3	$\pm$	0.2	\\
99.7th percentile	&	1.5	&	1.6	&	1.1	&	2.2	&	1.4	&	1.5	&	1.7	&	1.6	$\pm$	0.3	\\
Maximum	&	1.7	&	1.7	&	1.1	&	2.2	&	1.5	&	1.5	&	1.8	&	1.6	$\pm$	0.3	\\
	&		&		&		&		&		&		&		&					\\
\hline
\multicolumn{9}{|c|}{\textbf{Mass $\bm{M~(M_\oplus)}$ of the simulated metal-rich planets}} \\
\hline	
\textbf{}	&	\textit{N0}	&	\textit{T1}	&	\textit{T2}	& 	\textit{T3} & \textit{T4} & \textit{T5} & \textit{T6} & \textbf{Average}\\
\hline
Minimum 	&	0.006	&	0.006	&	0.006	&	0.004	&	0.006	&	0.01	&	0.07	&	0.02 $\pm$ 0.02	\\
Mean value	&	1.0	&	1.0	&	0.6	&	1.4	&	1.0	&	1.0	&	2.9	&	1.3	$\pm$	0.7	\\
68th percentile	&	1.3	&	1.1	&	1.0	&	1.1	&	1.4	&	1.4	&	3.5	&	1.6	$\pm$	0.8	\\
95th percentile	&	2.5	&	2.9	&	1.3	&	6.4	&	2.4	&	2.7	&	6.1	&	4	$\pm$	2	\\
99.7th percentile	&	5.4	&	7.0	&	1.5	&	23.0	&	4.9	&	4.9	&	7.6	&	8	$\pm$	7	\\
Maximum	&	8.0	&	8.0	&	1.9	&	23.5	&	5.2	&	5.2	&	11.5	&	9	$\pm$	7 \\
\hline													
\end{tabular}
\label{tab:results_compact}
\end{center}
\end{table*}

\end{appendix}

\clearpage

\bibliographystyle{plainnat}
\bibliography{Refs_Cambionietal}

\end{document}